\documentclass[fleqn,usenatbib]{mnras}

\usepackage{newtxtext,newtxmath}

\usepackage[T1]{fontenc}
\usepackage{ae,aecompl}

\usepackage{graphicx}	
\usepackage{amsmath}	
\usepackage{amssymb}	
\usepackage{color}
\usepackage{multirow}
\usepackage{pdflscape}
\hypersetup{pdfauthor={A. J. Stewart},
               pdftitle={On the optical counterparts of radio transients and variables},
               pdfkeywords={radio continuum: general, radio continuum: transients, stars: variables: general, stars: statistics, galaxies: statistics, quasars: general},
               bookmarksnumbered=true}

\title[Optical counterparts of radio transients]{On the optical counterparts of radio transients and variables}

\author[Stewart et al.]{A. J. Stewart,$^{1,2,3}$\thanks{email:adam.stewart@sydney.edu.au} T. Mu\~{n}oz-Darias,$^{4,5,2,3}$ R. P. Fender$^{2,3}$ and M. Pietka$^{2,3}$
\\
$^{1}$ Sydney Institute for Astronomy, School of Physics, The University of Sydney, NSW 2006, Australia\\
$^{2}$ Astrophysics, Department of Physics, University of Oxford, Keble Road, Oxford OX1 3RH, UK\\
$^{3}$ Physics and Astronomy, University of Southampton, Highfield, Southampton, SO17 1BJ, UK\\
$^{4}$ Instituto de Astrof\'isica de Canarias, 38200 La Laguna, Tenerife, Spain\\
$^{5}$ Departamento de astrof\'isica, Univ. de La Laguna, E-38206 La Laguna, Tenerife, Spain
}

\date{Accepted 2018 June 20. Received 2018 June 20; in original form 2017 September 29}

\pubyear{2018}

   \volume{{\rm in press}}
\begin{document}
\label{firstpage}
\pagerange{\pageref{firstpage}--\pageref{lastpage}}
\maketitle

\begin{abstract}
We investigate the relation between the radio ($F_{\mathrm{r}}$) and optical ($F_{\mathrm{o}}$) flux densities of a variety of classes of radio transients and variables, with the aim of analysing whether this information can be used, in the future, to classify such events. Using flux density values between 1--10~GHz and the optical bands $V$ and $R$, we build a sample with a total of 12,441 $F_{\mathrm{r}}$ and $F_{\mathrm{o}}$ measurements. The sample contains both Galactic objects, such as stellar sources and X-ray binaries, and extragalactic objects, such as gamma-ray bursts and quasars. By directly comparing the two parameters, it is already possible to distinguish between the Galactic and extragalactic populations. Although individual classes are harder to separate from the $F_{\mathrm{r}}$--$F_{\mathrm{o}}$ parameter space to a high accuracy, and can only provide approximations, the basic approach provides an already useful foundation to develop a more accurate classification technique. In addition, we illustrate how example objects from different classes move in the parameter space as they evolve over time, offering a feature that could be used to reduce the confusion between classes. A small, blind test of the classification performance is also undertaken using a catalogue of FIRST transient and variable sources, to demonstrate the advantages and current limitations of the approach. With more multi-wavelength data becoming available in the future, we discuss other classification techniques which the $F_{\mathrm{r}}$--$F_{\mathrm{o}}$ method could be combined with and potentially become an important part of an automatic radio transient classification system.
\end{abstract}



\section{Introduction}
\label{sec:intro}
The next generation of radio telescopes will survey the radio sky to unprecedented levels. This has the potential to uncover a wide variety of radio transient and variable phenomena, from both expected, and unexpected, origins. In recent years various large area surveys dedicated to searching for such radio sources have been carried out \citep[e.g.,][]{ofek2011,WilliamsSearch,Rowlinson,kunal} with some utilising the first wave of new instruments such as the upgraded Karl G. Jansky Very Large Array \citep[VLA;][]{vla}, and at low frequencies, the Low Frequency Array \citep[LOFAR;][]{lofar} and the Murchison Widefield Array \citep[MWA;][]{mwa}. These will shortly be joined by a group of new centimetre wavelength instruments such as MeerKAT \citep{meerkat}, ASKAP \citep{askap} and Apertif/WSRT \citep{apertif}. Lastly, the Square Kilometre Array \citep[SKA;][]{ska} in the 2020s, has the potential to uncover thousands of transients \citep{metzger, Fender}. At optical wavelengths, surveys such as the Palomar Transient Factory \citep[PTF;][]{ptf1,ptf2,ptf3} and those using the Pan-STARRS1 telescope \citep{panstarrs} have produced a high yield of transient sources \citep[e.g.][and the Pan-STARRS1 survey for transients currently lists over 10,000 transient sources since 2013\footnote{\url{https://star.pst.qub.ac.uk/ps1threepi/psdb/}}]{panstarrs-example,PTFexample}. The future Large Synoptic Survey Telescope \citep[LSST;][]{lsst} will view the optical sky to an unprecedented level of detail. The quality of images combined with a fast survey speed and a wide field of view could result in an estimated 10~million events per night \citep{lsstclassify2,lsst10mil}.

Our aim in this paper is to produce an initial analysis of the photometric properties of radio transients and variables along with their optical counterparts - providing a foundation to classify blindly discovered radio sources. In this work we consider radio transients that are found in the image-plane. These are usually incoherent synchrotron events that can be both Galactic, such as flare stars (though flare stars largely start to be dominated by coherent emission $\lesssim5$~GHz; \citealt{flarestarcoherent}), and extragalactic such as gamma-ray bursts and supernovae (we refer the reader to \citealt{metzger} and \citealt{Fender} for reviews of radio transient sources). We also consider radio pulsars - an example of coherent emission. Historically, a handful of radio transient events have been discovered through blind surveys \citep[e.g.][]{lev, galyam, ofek2010, bower, frail, NCP}, however, discoveries that are only detected in the radio make identifying the class of the transient object challenging. Although information can be gained from the nature of the radio emission itself, for example the time-scale of the event \citep{Pietka}, the use of simultaneous or rapid multi-wavelength follow up data becomes crucial in order to establish the type of the transient source in question. This was demonstrated by \citet{kunal} where the authors used various multi-wavelength follow-up, primarily optical in cohesion with PTF, to identify transient and variable sources detected in a 50~deg$^2$ pilot survey at 3~GHz with the VLA. On the optical side, we note that we primarily consider optical point-like sources since they account for the vast majority of the variable and transient populations and for the uniformity they offer. Though radio transient sources can appear behind extended optical objects which may offer clues as to their classification, but on the other hand, confirming the association between the two can be challenging.

The likely high number of radio transient events in the SKA era means that detailed follow-up of every event will be implausible. This means that rapid classification techniques will be necessary such as the ones widely utilised in optical surveys \citep[][]{Panstarrsclass, opticalclass2}. These commonly involve machine learning algorithms based upon light-curve features \citep{opticalref2} and other measurements such as optical colour \citep{opticalref1}. Automatic classification has also been investigated at X-ray wavelengths, e.g. \citet{Xrayclass} used random forest algorithms to define a classification of X-ray variable sources based upon features derived from time series, spectra and multi-wavelength catalogue data \citep[also see][]{Xrayclass2}. In the radio, attempts have been made to classify objects based upon light-curve features and the previously mentioned time-scale of the event \citep{radioclass,PietkaClassifier}. However, the need to rapidly classify objects has not been a requirement for radio transient astronomy, with efforts instead focussed on the rapid detection of events. We aim to provide the foundation that will lead to a similar system developed in the radio and optical parameter space.

Our selection of the radio--optical properties is due to the growing number of robotic optical telescopes such as the Liverpool Telescope \citep{LT} in the northern hemisphere, SMARTS\footnote{SMARTS is the Small and Medium Aperture Research Telescope System; see \url{http://www.astro.yale.edu/smarts/}.} in the south, and projects such as the Las~Cumbres Observatory Global Telescope Network \citep{LCOGT}. We expect rapid, or simultaneous, optical follow-up to be the most abundant for radio transients. This is in addition to present and future optical catalogues (e.g. the Sloan Digital Sky Survey \citep[SDSS;][]{SDSSIII} and LSST). For rapid classifications we also aim to only use basic properties of observations that are easily obtainable - the raw radio and optical flux density measurements of the observed object that would avoid expensive spectroscopic campaigns for a preliminary classification. A project that is very applicable to the above scenario is the MeerLICHT telescope\footnote{MeerLICHT is a prototype telescope for the BlackGEM telescope array; see \url{http://www.ast.uct.ac.za/meerlicht/MeerLICHT.html}.}. It is a robotically operated 65~cm telescope with a 2~deg$^2$ field of view that will follow MeerKAT observations and provide simultaneous optical coverage. 

We first construct a sample of radio and optical flux density values of known radio transient and variable objects, which is described in Section~\ref{sec:method}. This is followed by Section~\ref{sec:results} which shows the results that are discussed in Section~\ref{sec:discussion} along with the exploration of the transient diagnostic application of the data. We  finish with our conclusions in Section~\ref{sec:conclusions}.

\section{Data collection \& methods}
\label{sec:method}
We collected radio and optical flux density measurements of various types of transient objects from the literature. We define these as $F_{\mathrm{r}}$ and $F_{\mathrm{o}}$ respectively throughout this paper. The classes of object for which data were gathered were as follows: quasar, stellar, radio pulsar, cataclysmic variable (CV), X-ray binary (XRB), gamma-ray burst (GRB) and supernova (SN). We only consider detections of objects at both wavelengths and currently include no upper limit values. In cases where no tabled observational data was available, measurements were obtained from light-curve figures using {\sc WebPlotDigitizer}\footnote{Web based tool to extract data from plots, images and maps, authored by Ankit Rohatgi (\url{http://arohatgi.info/WebPlotDigitizer}).}. We also note that because of how the sample is constructed, i.e. from many different sources, the sample will inevitably be very biased. At this early stage of the project we do not attempt to un-bias the dataset due to the complexity of the task. We discuss such issues in Section~\ref{sec:discussion}.

\subsection{Measurement selection criteria}
Requiring many different classes of objects meant collecting data from a range of projects, all done with different methods, instruments and data quality. Hence, we decided upon following a selection criteria to provide a useful sample in terms of size and quality. In the optical case we concentrated on gathering $V$ and/or $R$ band measurements. The two bands were found to be the most common, and for our purposes were considered valid to be used interchangeably. If both were available then $R$ was used and if only an SDSS-$r$ band measurement was available for a source, we recorded it as $R$ - we assume that the difference is negligible compared to the scatter created by using both $V$ and $R$. Nearly all optical magnitudes obtained were recorded in the Johnson-Cousins $BVRI$ system, with the exception of optically selected quasars which were gathered from the SDSS survey in the $r$~band, which is considered to be in the AB system, four GRBs and one SN. As the quasars are by far the dominant class and the preference to work in the AB system, we account for this by converting all Johnson-Cousins magnitudes to the AB system using the offsets as defined by \citet{ABconversion} (specifically the author's table~2). Any sources that were collected in the SDSS-$r$ filter were not converted. We also do not account for extinction on optical magnitudes to mimic values obtained directly from the telescope.

The $F_{\mathrm{r}}$ measurements selected were in the frequency range of 1--10~GHz. Using a large portion of the radio bandwidth meant that the dataset would be useful for a range of common radio frequencies - given the overall aim of providing a classification to radio sources. The large range also meant that a sizeable sample could be created. We believe the range of frequencies does not have a negative effect on our initial investigation as discussed further in Section~\ref{sec:discussion}.

In addition, all the populations can experience rapid flux density variations, hence, the timing between the $F_{\mathrm{r}}$ and $F_{\mathrm{o}}$ measurements was also important. Ideally, simultaneous measurements are required, with our definition of `simultaneous' for the purposes of this work meaning recorded $\leq1$~d apart. However, while possible with some classes, the data did not exist for others. This is particularly true for quasars and stellar sources due to the use of existing catalogues to obtain the measurements. Simultaneous measurements for pulsars were also found to be too rare, so non-simultaneous measurements were used. 

There is also the issue of whether the objects are in a quiescent or transient/flaring state when the flux density measurements are recorded. For example, classes such as GRBs and SN are always in a `transient' state, where as stellar and quasar sources can be seen as `quiescent' that can display flaring behaviour. This will create extra biases in our sample, e.g. some classes may be dominated by flaring objects as this is what has originally triggered the observations of the source. Though sources in a flaring state are also those more likely to be discovered as a `new source' transient object, which are the primary focus of this work. Throughout the description of the sample we clarify what states we believe the sources to be in, or give the exact nature where possible. The two classes where catalogues have been used, the quasars and stellar sources, we assume to be dominated by sources in a quiescent state, as neither catalogue had any selection criteria suggesting otherwise (see the following Sections~\ref{sec:quasars} and \ref{sec:stellar}). Given that these classes are dominant objects in the sky, we somewhat consider them as a measure of the `background'. We also attempt to be consistent in our selections of what enters the final result for other classes (e.g. using radio peak measurements for flaring objects) and we also address this issue by exploring how transient objects evolve over time in the $F_{\mathrm{r}}$ and $F_{\mathrm{o}}$ domain (Section~\ref{sec:dynamicdata}).

The following sections give a summary of data collected for each class of object, with Table~\ref{table:totals} showing the final total number of measurements gathered for each class. Where applicable, the references for the individual measurements can be found in Appendix~\ref{app:tables} along with distance measurements, and in the cases of SN and GRB, the age of the transient when the used measurements were recorded. Fig.~\ref{fig:histograms} shows the distributions of the radio frequencies and optical bands used for each class in the sample. In addition, the redshift, or distance, distributions for the large samples of quasars, stellar sources and GRBs are also shown in Figs.~\ref{fig:qsozhistogram}, \ref{fig:stellardisthistogram} and \ref{fig:grbzhistogram} respectively. 

\begin{table}
	\centering
	\caption{The total number of pairs of $F_{\mathrm{r}}$ and $F_{\mathrm{o}}$ measurements included in the sample for each class of object. Also shown is the total number of optical colour measurements which were obtained where available as an additional reference for our sample. The potential use of colour is discussed in Section~\ref{sec:discussion}. }
	\label{table:totals}
	\begin{tabular}{cc} 
		\hline
		Object Class & Total No.\\
		\hline
		Quasar (Optical Sel.) & 11,062\\
		Quasar (Radio Sel.) & 720 \\
		Stellar & 534 \\
		Radio pulsar & 7 \\
		CV & 18 \\
		XRB & 26 \\
		GRB & 48 \\
		SN & 26 \\
		\hline
		Total & 12,441\\
		\hline
	\end{tabular}
\end{table}

\subsection{Quasars}
\label{sec:quasars}
We used two different selection methods to populate our quasar sample. First, we selected quasars which had been classified as such through their optical properties, and second, a sample of quasars which had been defined by their radio properties.

\subsubsection{Optically selected}
\label{sec:opticalquasars}
Two versions of the SDSS quasar catalogue were used: the Seventh Data Release quasar catalogue \citep[DR7QC;][]{SDSSQCDR7} and the Ninth Data Release quasar catalogue \citep[DR9QC;][]{SDSSQCDR9}. The DR7QC consisted of data from the SDSS-I and SDSS-II surveys \citep{SDSS} and contains 105,783 entries, where as the DR9QC was compiled using SDSS-III \citep{SDSSIII} and contains 87,822 quasars. The two catalogues contain cross-matches to the VLA FIRST 1.4~GHz survey \citep[resolution of $5^{\prime\prime}.4$;][]{FIRST} and record the peak $F_{\mathrm{r}}$. These are matches within $2.0^{\prime\prime}$ of the SDSS quasar position, and are likely to be cores of the galaxy. If any quasars were present in both the SDSS catalogues, then only the information from the DR9QC was used. In total, 11,062 optically selected quasars are included in our sample.

Combining the two SDSS quasar catalogues provided us with a diverse range of objects, particularly in redshift, $z$, as this is the main difference between the two. The DR9QC was built using data from the Baryon Oscillation Spectroscopic Survey of SDSS-III \citep[BOSS;][]{BOSS}, which targeted over 150,000 $z>2.15$ quasars. This resulted in DR9QC containing 2.6 times more quasars at $z > 2.15$ than the DR7QC, for which the majority of quasars were at $z<2.15$ (see Fig.~\ref{fig:qsozhistogram} for the QSO sample redshift distribution). We note that some of the optical quasars may have been originally identified as such by the presence of a radio counterpart, but for our purposes we define them as `optically selected'.

We use the $r$ band magnitude from the SDSS survey which we assume to be already in the AB magnitude system hence no conversion is applied. We also note that the 1~mJy limit of the FIRST survey used in this class is a significant limiting factor for the project, however it remains the best source of defining the radio properties of the quasar population. The consequences of the limit are discussed further in Sections~\ref{sec:results}~\&~\ref{sec:discussion} along with how we would expect the population to appear beyond the 1~mJy limit.

\subsubsection{Radio selected}
\label{sec:radioquasars}
The basis of our radio selected sample was the Parkes Catalogue \citep{PKS}, where we selected any object identified as a quasar or a BL~Lacertae. For each object, the $F_{\mathrm{r}}$ at 5~GHz was obtained (1.4~GHz data was poorly sampled), and using SIMBAD, we gathered $V$ and/or $R$ photometry measurements when available. Assuming a resolution of 3.3 arcmin for Parkes ($D=64$~m) at 5~GHz, we assume the total $F_{\mathrm{r}}$ is recorded. In total, 720 radio selected quasars were included in the sample.

\subsection{Stellar objects}
\label{sec:stellar}
To characterise the Galactic stellar sources, we have used a catalogue of $F_{\mathrm{o}}$ and $F_{\mathrm{r}}$ properties as used by \citet{Gudel} (in particular the information presented in their fig.~1). The radio measurements of this catalogue are in turn based upon stellar radio detections between 1--10~GHz, as reported by \citet{Wendker}. $F_{\mathrm{o}}$ measurements in the $V$ band were also obtained by \citet{Wendker} from a wide range of sources in the literature at the time. The classes of objects stated in the catalogue were as follows: RS CVn and Algol-type stars; double stars; magnetic stars; BY Dra-type stars; Symbiotic stars; Herbig stars; WR stars; T Tauri stars; PMS stars; and isolated stars. We checked the classifications against the most recent literature and removed any source that were now classed as: X-ray binary; X-ray source; X-ray nova; planetary nebular; QSO; and radio galaxy. We further generalised the subcategories into: Stars, RS~CVn, Algol, Young Stellar Objects (YSOs), Variable Stars, Symbiotic Stars and `other'. CVs were also removed and integrated into the stand-alone CV data (refer to Section~\ref{sec:CVs}). We note that while we have defined CVs and X-ray binaries as standalone samples, it will be shown in Section~\ref{sec:results} that using all the sub-classifications of the stellar sample was not useful.

In total, 534 stellar sources are included. Similar to the quasar sample, using a catalogue approach meant that we can sample all stellar objects in a variety of astrophysical states. We assume this to be true given that no strict selection criteria was imposed by \citet{Wendker}, in which optical magnitudes were gathered from a wide range of literature sources.

\subsection{Isolated neutron stars}
\label{sec:neutron}
\subsubsection{Radio pulsars}
We have included seven radio pulsars in the sample: Crab~pulsar, Vela~pulsar, B0540-69, B1509-58, B0656+14, B1133+16 and B0950+08. As simultaneous observations are rare, we have used optical observations of the pulsars performed by \citet{pulsars_optical}, paired with radio observations that were compiled in the ATNF Pulsar Catalogue \citep{pulsar_cat}. The radio measurements are all at 1.4~GHz. Reference details can be found in Table~\ref{table:pulsars}.

\subsubsection{Other isolated neutron stars}
We also attempted to gather measurements for other types of isolated neutron stars such as magnetars and RRATs. As of writing, the McGill Online Magnetar Catalogue \citep{McGill}\footnote{\url{http://www.physics.mcgill.ca/~pulsar/magnetar/main.html}.} lists four magnetars  with radio observations. However, these could not be matched with optical detections and thus were not used. The same is true with RRATs where no optical detections have been made \citep{RRATlimits1,RRATlimits2}.

\subsection{Cataclysmic variables}
\label{sec:CVs}
Gathering a sample of CVs in the context of our project is challenging given the dynamic nature of these objects, which can undergo large variations in brightness over relatively short periods of time. While CVs are commonly persistent sources that undergo flaring events, they are likely to be discovered as new transient events when such a flare is detected above the detection limits. Hence, CVs (along with other classes such as XRBs, SN and GRBs) will significantly move around the $F_{\mathrm{o}}$--$F_{\mathrm{r}}$ parameter space, a feature that we explore in this work by gathering a dynamic data sample (Section~\ref{sec:dynamicdata}).

In total, we have included 17 CVs from a range of sources in a conscious attempt to sample the class during quiescence and flaring events. Of these, 7 are from the stellar sample as discussed in Section~\ref{sec:stellar}. These are T~CrB; V*~RT~Ser; RS~Oph; V~4074~Sgr; HM~Sge; PU~Vul and AG~Peg. All are nova type CVs, and the precise timing between the optical and radio observations is unknown, hence we treat these as quiescent measurements (though this is not certain). The 10 other CVs were gathered as follows. Five nova CVs (Nova~Sco~2012, T~Pyx, V1500~Cyg, V1723~Cyg and V1974~Cyg) were included from the work performed by \citet{Pietka} on studying radio light-curve properties of objects, and are all flaring events. Three novalike CVs that were detected by \citet{Deanne} in the radio at 6~GHz (RW~Sex, V603~Aql and TT~Ari) that are considered as being in quiescence. Although, one measurement of TT~Ari showed a radio flare in one of the author's observations, and hence, has two measurements included in the sample. Lastly, two dwarf~nova, SS~Cyg and V3885~Sgr, where the former is recorded at the radio peak of an outburst, and the latter is a targeted detection and is considered in quiescence. All these measurements are in the frequency range 2.7--8.6~GHz and the known distances for the objects included range 0.12--5.0~kpc. For the sources outside of the stellar sample, if no published $F_{\mathrm{o}}$ measurements were available, the $V$ or $R$ band magnitudes closest to the date of the radio observations, were obtained from AAVSO\footnote{American Association of Variable Star Observers; see \url{https://www.aavso.org}.} light-curves. The measurements of the $F_{\mathrm{o}}$ and $F_{\mathrm{r}}$ of the sources from \citet{Pietka} and those of SS~Cyg are within 1~d of each other in time. However the V3885~Sgr measurements are not simultaneous, and are separated by 38~d. The time separation between the $F_{\mathrm{o}}$ and $F_{\mathrm{r}}$ measurements of the \citet{Deanne} sources was 1~yr, 1~d and 22~d for RW~Sex, V603~Aql and TT~Ar, respectively. In total, 18 pairs of measurements were used for CVs. Reference details can be found in Table~\ref{table:CVs}. Five CVs have been included in the dynamic data sample, of which the details can be found in Table~\ref{table:dynamic} and is discussed in Section~\ref{sec:dynamicdata}.

\subsection{X-ray binaries}
\label{sec:x-raybs}
Similar to CVs above, the brightness of XRBs can vary by large magnitudes during flare events and numerous measurements are made (though not all) on the basis of an X-ray transient detection. To build the sample we used two catalogues: the Catalogue of high-mass X-ray binaries in the Galaxy \citep[4$^{\textrm{th}}$~edition;][]{HMXBcat} (LiuHMXB) and the Catalogue of low-mass X-ray binaries in the Galaxy, LMC, and SMC \citep[4$^{\textrm{th}}$~edition;][]{LMXBcat} (LiuLMXB). The XRBs with a high confidence radio detection as defined in the catalogues (doubtful radio detections were excluded) along with an optical detection were considered, which meant referring to the literature on the object to determine if there were suitable data. Three types of measurements were used in the sample: (i) radio and optical observations following the discovery of the respective object as an X-ray transient. These observations are commonly performed within days to weeks after the discovery and in this case we record the pair of measurements at the radio peak if possible. (ii) targeted (often in the radio) observations of known XRBs in an attempt to gain a detection, and (iii) long term monitoring (usually at different points in time for radio and optical) for which an average flux density can be recorded. If major flaring occured during the monitoring of the respective source then we did not include the object. Incorporating these different methods also had the advantage of providing a range of measurements over the $F_{\mathrm{r}}$--$F_{\mathrm{o}}$ parameter space.

In total, 26~XRB objects were included: 6 from LiuHMXB and 20 from LiuLMXB. The majority of the measurements (19) were within 1~d of each other, with a further four up to 4~d. The final three were 1~week~(2) and $\sim$20~yr apart (the latter being an average flux density). Reference details for each XRB can be found in Table~\ref{table:x-ray}, along with notes of the circumstance of the used observations. These show that the sample is dominated by flaring events at the radio peak as the majority are follow-up observations as described above. The radio measurements span 1--9~GHz and the distance of the objects span 1--27~kpc . Eight of the XRBs were also included in the dynamic data sample of which details can be found in Table~\ref{table:dynamic} and is discussed in Section~\ref{sec:dynamicdata}.

\subsection{Supernovae \& Gamma-ray bursts}
\label{sec:SNandGRB}
Our approach for SNe and GRBs was to use the first simultaneous ($\leq1$~d) measurements available in the literature. Note that this is not necessarily the first detection measurement available at either wavelength, as these were unlikely to be simultaneous. All the $F_{\mathrm{r}}$ and $F_{\mathrm{o}}$ measurements included in both samples are $\leq1$~d apart in observation time, with the sample size large enough to avoid including non-simultaneous measurements.

\subsubsection{Supernovae}
\label{sec:SNe}
We have used a sample of radio detected core-collapse SNe compiled by \citet{Romero}. The sample includes a variety of SNe types: Ib, Ic, II, IIn, IIb, IIP and a full list of the 26 SNe can be found in Table~\ref{table:SNe} along with references and distances. The measurements range from 3--743 d after the reported explosion dates, with a distance range of 3.4--154.0~Mpc (see Fig.~\ref{fig:stellardisthistogram}) and are within the radio frequency range of 4.8--8.6~GHz. From the sample, 15 SNe were also included in the dynamic data sample of which details can be found in Table~\ref{table:dynamic} and is discussed in Section~\ref{sec:dynamicdata}. The SN sample was compiled with the aid of the valuable resource \textit{The Open Supernova Catalog}~\citep{OpenSNCatalog}.
\begin{table*}
	\centering
	\caption{The information regarding the sample of dynamic sources we have used in the investigation. The first and last dates of the measurements refer to the $F_{\mathrm{r}}$ measurement. `Timespan' refers to the total number of days between the first pair of $F_{\mathrm{r}}$ and $F_{\mathrm{o}}$ measurements and the last. `No. of data points' refers to the number of simultaneous pairs of $F_{\mathrm{r}}$ and $F_{\mathrm{o}}$ that were obtained. For further details on the specifics of the sample, such as when measurements began relative to the event being detected, please refer to Section~\ref{sec:dynamicdata} and Tables~\ref{table:CVs}, \ref{table:x-ray}, \ref{table:SNe} and \ref{table:GRBs}. The references are as follows: [1]~\citet{RSOph_dynamic_r}, [2]~AAVSO, [3]~\citet{SSCyg}, [4]~\citet{TPyx}, [5]~\citet{V1500Cyg_r}, [6]~\citet{V1974Cyg_r}, [7]~\citet{xtej0421}, [8]~\citet{groj0422}, [9]~\citet{J165540_r_early}, [10]~\citet{groj1655_r}, [11]~SMARTS \citep[see][]{Buxton}, [12]~\citet{grs1124_r}, [13]~\citet{grs1124_o}, [14]~\citet{Corbel}, [15]~\citet{Buxton}, [16]~\citet{Han}, [17]~\citet{Casares}, [18]~\citet{xtej1550_r}, [19]~\citet{xtej1550_o}, [20]~\citet{xtej1859_r} (GBI), [21]~\citet{xtej1859_o}, [22]~\citet{vanDyk93}, [23]~\citet{Clocchiatti}, [24]~\citet{Weiler}, [25]~\citet{Richmond}, [26]~\citet{SN1994I_r}, [27]~\citet{SN1994I_o}, [28]~\citet{SN1994I_tracko2}, [29]~\citet{SN1994I_tracko3}, [30]~\citet{SN1998bw_r}, [31]~\citet{SN1998bw_track}, [32]~\citet{Berger}, [33]~\citet{Gal-Yam}, [34]~\citet{2004dj_r_dynamic}, [35]~\citet{2004dj_o_dynamic}, [36]~\citet{Wellons}, [37]~\citet{Drout}, [38]~\citet{2004et_r_dynamic}, [39]~\citet{2004et_o_dynamic}, [40]~\citet{SN2007bg_r}, [41]~\citet{SN2007bg_o}, [42]~\citet{SN2007gr_r}, [43]~\citet{SN2007gr_o}, [44]~\citet{SN2007uy}, [45]~\citet{SN2008D_r}, [46]~\citet{SN2008D_o}, [47]~\citet{Roming}, [48]~\citet{SN2009bb_r}, [49]~\citet{SN2009bb_o}, [50]~\citet{GRB970508_r}, [51]~\citet{GRB970508_o}, [52]~\citet{GRB991208_dynamic_r}, [53]~\citet{GRB991208_o}, [54]~\citet{GRB000301C_r}, [55]~\citet{GRB000301C_dynamic_o}, [56]~\citet{GRB030329_r}, [57]~\citet{GRB030329_o}, [58]~\citet{GRB050820A}, [59]~\citet{GRB060218_r}, [60]~\citet{GRB060218_o}, [61]~\citet{GRB070125_dynamic}, [62]~\citet{GRB100418A_r}, [63]~\citet{GRB100418A_track}, [64]~WEBT Archive, \citet{BLLac} and [65]~GASP-WEBT archive, \citet{3C454_ro}. WEBT is the Whole Earth Blazar Telescope and GASP the GLAST-AGILE Support Program.}
	\label{table:dynamic}
	\resizebox{\textwidth}{!}{\begin{tabular}{cccccccccc} 
		\hline
		Object & Object & Date of first & Date of last & Timespan & No. of & Radio freq. & Optical & Radio & Optical\\
		name& class & flux densities & flux densities & (d) & data points & (GHz) & band & ref. & ref. \\
		\hline
		RS Oph & CV, Nova & 2006 Feb 18& 2006 Apr 03& 45 & 9 & 6.0 & $V$ & [1] & [2]\\
		SS Cyg &CV, Dwarf Nova &2007 Apr 25 &2007 May 10 &15 & 8 & 8.6 & $V$ & [3] & [3]\\
		T Pyx &CV, Nova & 2011 June 19&2012 Sept 07 & 445& 10 & 5.0 & $V$ & [4] & [4]\\
		V1500 Cyg &CV, Nova &1975 Sept 24 &1976 Oct 01 &373 & 11 & 8.1 & \textit{Vis.} & [5] & [2] \\
		V1974 Cyg &CV, Nova &1992 May 04 &1994 Sept 14 & 862& 17 & 5.0 & \textit{Vis.} & [6] & [2]\\
		\hline
		XTE J0421+560 & HMXB & 1998 Apr 04 & 1998 Apr 21 &17 & 11 & 8.0 & $R$ & [7] & [7] \\
		GRO J0422+32 & LMXB & 1992 Aug 13 & 1993 Jan 15 & 155 & 6 & 5.0 & $V$ & [8] & [8]\\
		GRO~J1655-40 & LMXB & 2005 Feb 20 & 2005 Apr 05 & 45 & 16 & 4.86 & $V$ & [9,10] & [11] \\
		GS 1124-684 & LMXB & 1991 Jan 18 & 1991 Jan 31 & 13 & 8 & 4.7 & $R$ & [12] & [13]\\
		GX~339-4 & LMXB & 2011 Feb 06 & 2011 Apr 27 & 80 & 15 & 9.0 & $V$ & [14] & [15] \\
		V404~Cygni & LMXB & 1989 May 30 & 1989 July 04 & 35 & 7 & 8.3 & $R$ &  [16] & [17] \\
		XTE J1550-564 & LMXB & 1998 Sept 20 & 1998 Sept 29 & 9 & 7 & 4.8 & $R$ & [18]&[19] \\
		XTE J1859+226 & LMXB & 1999 Oct 17 & 1999 Dec 31 & 74 & 46 & 2.25 & $R$ &[20] &[21] \\
		\hline
		SN 1990B & SN, Type Ic& 1990 Feb 13 & 1990 May 29 & 105 & 5 & 5.0 & $V$ &[22] &[23] \\
		SN~1993J & SN, Type II & 1993 Apr 08 & 1994 Feb 08 &306 & 34 & 8.3 & $R$ & [24] & [25] \\
		SN~1994I & SN, Type Ic & 1994 Apr 03 & 1994 Aug 08 &127 & 26 & 8.3 & $R$ &  [26] & [27,28,29] \\
		SN~1998bw & SN, Type Ic & 1998 Apr 28 & 1998 July 15 & 78& 22 & 8.64 & $R$ & [30] & [31] \\
		SN~2002ap & SN, Type Ic & 2002 Feb 01 & 2002 Feb 18 & 17 & 7 & 1.43 & $R$ &[32] &[33] \\
		SN~2004dj & SN, Type IIP& 2004 Aug 05 & 2004 Nov 24 & 110 & 11 & 5.0 & $V$ &[34] &[35] \\
		SN 2004dk & SN, Type Ib& 2004 Aug 07 & 2004 Sept 18 & 41 & 5 & 8.5 &$R$ &[36] & [37] \\
		SN 2004et & SN, Type II& 2004 Oct 07& 2004 Dec 03 & 57 & 9 & 5.0 &$R$ & [38] & [39] \\
		SN 2004gq  & SN, Type Ib& 2004 Dec 08 & 2005 Jan 21 & 35 & 5 & 8.5 & $R$ & [36] &[37] \\
		SN 2007bg & SN, Type Ic& 2007 Apr 30 & 2007 Jul 24 & 68 & 6 & 8.46 & $R$ & [40] &[41] \\
		SN 2007gr & SN, Type Ic&2007 Aug 17 &2007 Nov 18 & 92 & 8 & 4.9 & $R$ & [42] & [43] \\
		SN 2007uy & SN, Type Ib& 2008 Jan 11 & 2008 Mar 07 & 56 & 12 & 8.4 & $R$ &[44] &[44] \\
		SN 2008D & SN, Type Ib& 2008 Jan 15 & 2008 May 10 & 115 & 7 & 4.8 & $R$ & [45] &[46] \\
		SN 2008ax & SN, Type IIb& 2009 Mar 07 & 2008 Apr 21 & 46 & 9 & 8.46 & $V$ &[47] &[47] \\
		SN 2009bb & SN, Type Ic& 2009 Apr 05 & 2009 May 20 & 44& 10 & 8.46 & $V$ &[48] &[49] \\
		\hline
		GRB 970508 & GRB & 1997 May 15 & 1997 Aug 04 & 82 & 11 & 8.46 &$R$ &[50] &[51] \\
		GRB 991208 & GRB & 1999 Dec 10 & 2000 Jan 17 & 37 & 7 & 8.46 &$R$ &[52] &[53] \\
		GRB 000301C & GRB & 2000 Mar 05 & 2000 Apr 18 & 43 & 6 & 8.46 &$R$ &[54] & [55]\\
		GRB~030329 & GRB & 2003 Mar 30 & 2003 June 04 & 65& 20 & 8.64 & $R$ & [56] & [57] \\
		GRB~050820A & GRB & 2005 Aug 20 & 2005 Sept 15/26$^*$ & 25 & 6 & 8.46 & $R$ & [58] & [58] \\
		GRB 060218 & GRB & 2006 Feb 20 & 2006 Mar 15 & 23 & 12 & 8.46 &$R$ & [59] & [60] \\
		GRB 070125 & GRB & 2007 Jan 31 & 2007 Feb 18 & 18 & 6 & 8.46 &$R$ & [61] & [61] \\
		GRB~100418A & GRB & 2010 Apr 20 & 2010 May 21 &31 & 7 & 8.46 & \textit{white filter} & [62] & [63] \\
		\hline
		BL~Lacertae & Quasar & 1994 June 28 & 2005 Jan 12 & 3851 & 265 & 5.0 & R & [64] & [64] \\
		3C~454.3 & Quasar & 2008 May 08 & 2010 Jan 28 & 630 & 79 & 8.0 & R & [65] & [65] \\
		\hline
	\end{tabular}}
		\begin{flushleft}
	$^*$ In the case of GRB~050820A the last pair of $F_{\mathrm{o}}$ and $F_{\mathrm{r}}$ measurements were separated by 11~d.\\
	\end{flushleft}
\end{table*}

\begin{figure*}
	\includegraphics[width=0.995\textwidth]{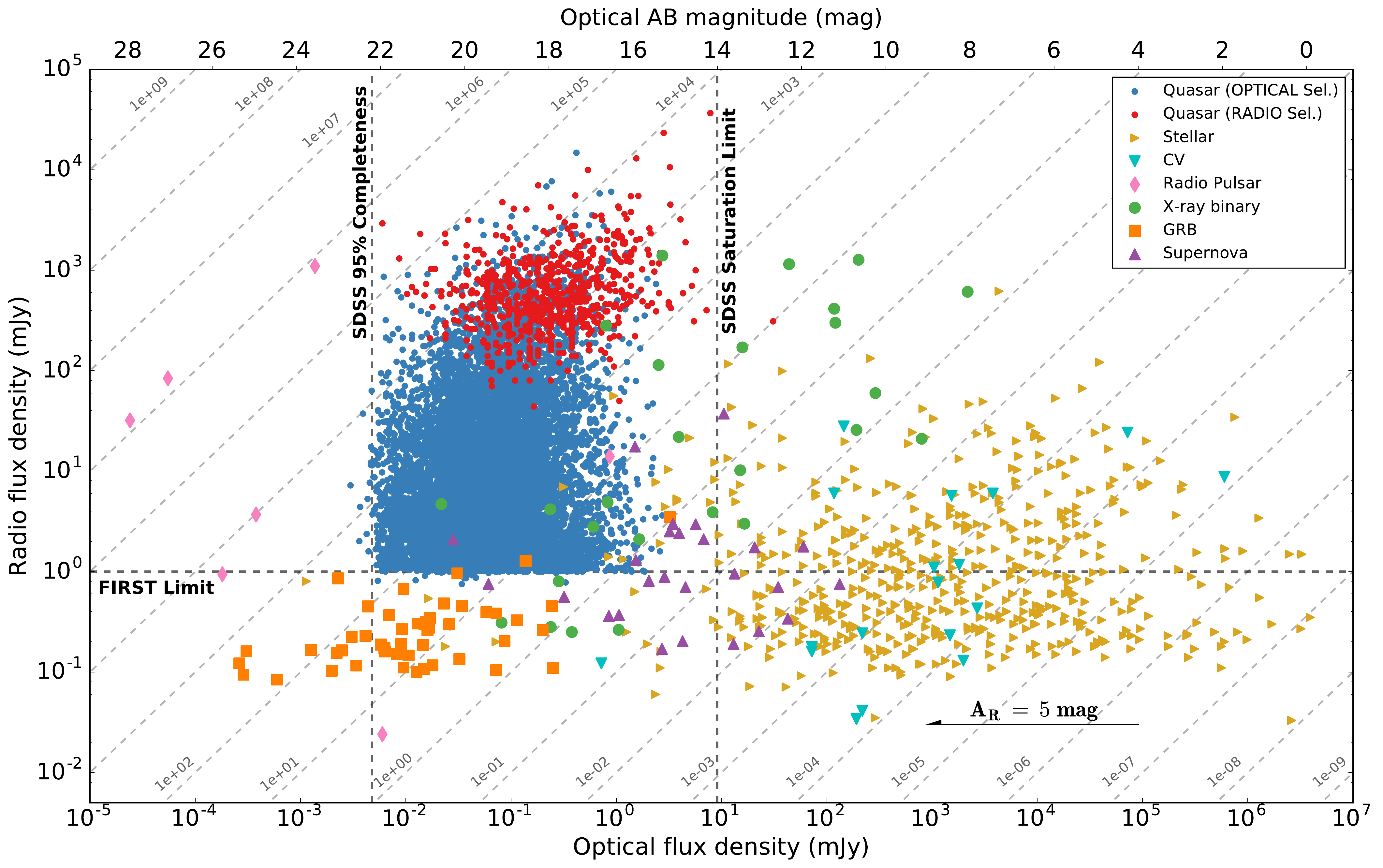}
    \caption{$F_{\mathrm{o}}$ versus $F_{\mathrm{r}}$ of our sample of objects from different classes. The $F_{\mathrm{o}}$ values have been converted from mag to mJy, with the optical magnitude scale given as a reference on the top axis (based upon the $R$ band). The data contains $F_{\mathrm{o}}$ in bands $V$ and $R$ and the $F_{\mathrm{r}}$ covers a range of frequencies from 1--10~GHz. Labelled on the plot are: a horizontal line denoting the FIRST survey source detection threshold (1~mJy); a vertical line denoting the SDSS 95 per cent completeness limit ($\sim$22~mag), and another showing the SDSS saturation limit (14~mag). The diagonal lines in the background of the plot show the lines of constant ratio between the two flux density density values corresponding to changing distance and are labelled with respective ratio. A 5~mag scale arrow is also shown to show the possible effect of extinction on the optical magnitudes. Please refer to the online print for a colour version of the figure.}
    \label{fig:basis}
\end{figure*}
\subsubsection{Gamma-ray bursts}
\label{sec:GRB}
Our GRB sample is based upon a catalogue of radio GRB afterglows that was compiled by \citet{Chandra}. We only used GRBs that had simultaneous $F_{\mathrm{o}}$ and $F_{\mathrm{r}}$ measurements available. In total we have recorded measurements for 48~GRBs, with details presented in Table~\ref{table:GRBs}. Two of the bursts were short GRBs (defined as SHBs by \citealt{Chandra}), which were GRB~050724 and GRB~051221A, with the rest characterised as long GRBs. The measurements range from $<1$--26~d post burst, and the sample spans a redshift range of 0.03--4.50 (see Fig.~\ref{fig:grbzhistogram}) and are within the radio frequency range of 1.4--9.0~GHz. Eight GRBs were also included in the dynamic data sample of which details can be found in Table~\ref{table:dynamic} and is discussed in the following Section~\ref{sec:dynamicdata}.

\subsection{Dynamic Data}
\label{sec:dynamicdata}
The $F_{\mathrm{r}}$ and $F_{\mathrm{o}}$ of objects will change over time. This is an important feature to consider since the property could help to reduce confusion between classes when they overlap in the $F_{\mathrm{r}}$--$F_{\mathrm{o}}$ parameter space. In addition, the movement also helps to address the flaring or quiescent bias in the sample by showing an object's full range as opposed to a data point at the peak of a flare. A detailed study of how classes of objects evolve in the $F_{\mathrm{r}}$--$F_{\mathrm{o}}$ parameter space is beyond the scope of this initial investigation. Nevertheless, we gathered a dynamic data sample based upon our main sample to explore what could be expected. The sources selected were not dependant on any prior set conditions. We judged each object as the main sample was being collected and used those which had long-term (ideally $\geq1$~month) well sampled light-curves. Simultaneous measurements were still important although we relaxed the time required between measurements to 3~d in order to achieve a good sample size with well sampled light-curves. The sample, of which details can be found in Table~\ref{table:dynamic}, contains 5 transient/flare CV events, 8 XRB transient/flare events, 15 SNe, 8 GRBs and 2 quasars for a total of 38 dynamic sources. As our quasar sample did not allow for individual light-curves, two quasar objects with good coverage were specifically found to include. One of the quasars is BL~Lacerate which we consider a `quasar' for our purpose (as we did with the radio selected quasars). The coverage of these events range from 2 weeks to 10.5~yr. The only exception of the 3~d pair requirement is the last measurement of GRB~050820A, which were recorded 11~d apart due to lack of simultaneous coverage. Also, three sources have the $F_{\mathrm{o}}$ recorded in `Visible' (V1500~Cyg and V1974~Cyg; both CVs) or `white filter' (GRB~100418A) as these were the only optical bands available and were judged to be suitable for the dynamic purpose.

The `state' of the CVs and XRBs in the dynamic sample is not as trivial as for the SNe and GRBs, hence the details of the circumstances of these sources can be found in Tables~\ref{table:CVs} and \ref{table:x-ray}. In summary, the five CVs are measurements of outbursts where the first simultaneous measurements occur between 1.3--66~d after the optical maximum or outburst itself.
For the XRBs, seven are simultaneous measurements beginning between 3--14~d following the detection of an outburst at X-ray wavelengths. The exception is GX-339 where the measurements monitor the decline of a flare that occurred approximately 1~yr prior. During the decline the source undergoes a small re-flaring in both $F_{\mathrm{r}}$ and $F_{\mathrm{o}}$ as it transitions from the soft to hard state.

The radio frequencies and optical bands used have been kept as consistent as possible over the dynamic data. In addition, within the coverage of each source the entire light-curve is in one single radio/optical frequency/band only. We note that such long coverage of objects may signify a unique characteristic of the event. For example, SN~1993J and GRB~030329 represent some of the brightest radio events of their classes. Thus, while these sources offer a good example of how classes are likely to evolve, they are not fully representative.

 \begin{figure}
	\includegraphics[width=\columnwidth]{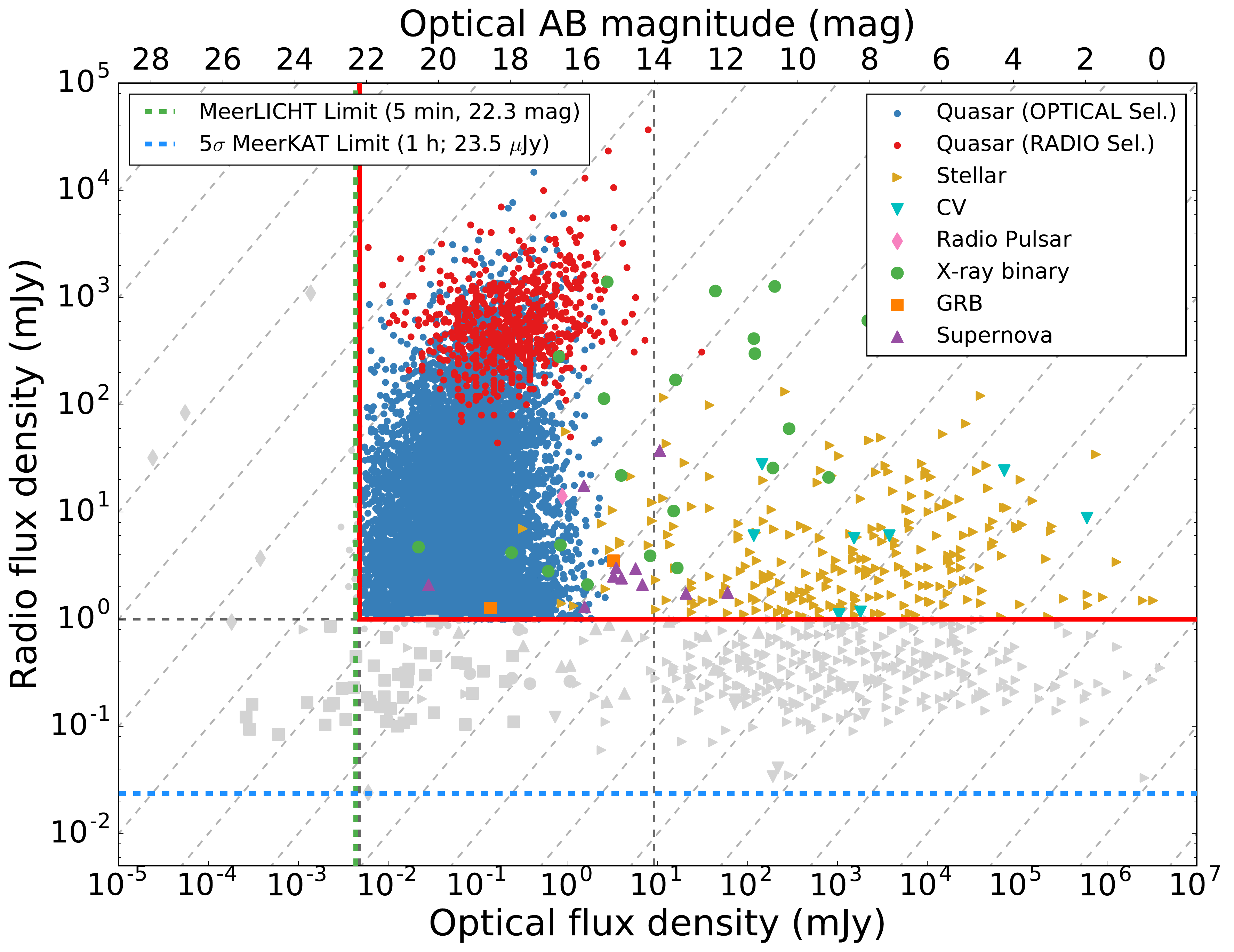}
    \caption{Denoted by solid red lines is the boundary for which our sample is the most `complete', i.e. those sources (plotted in colour) which are brighter than a $R$ magnitude of 22 in the optical (the SDSS 95\% completeness) and 1~mJy in the radio (the sensitivity limit of the FIRST survey). Also plotted are the limits that define the region that will be explored by the MeerKAT and MeerLICHT telescopes in the radio and optical respectively.}
    \label{fig:split}
\end{figure}

\begin{figure}
	\includegraphics[width=\columnwidth]{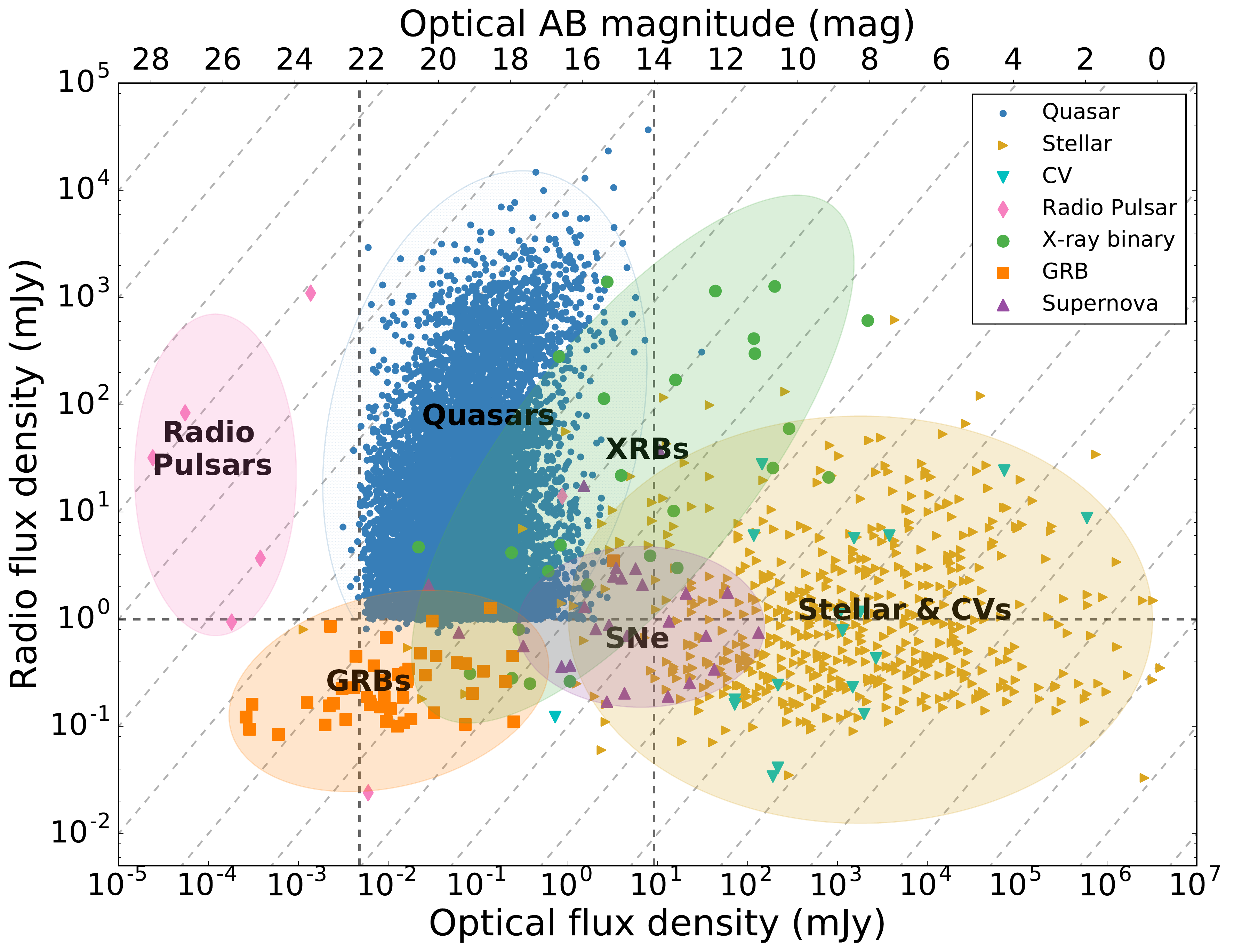}
    \caption{A simplistic schematic version of the base result where coloured ellipses represent the space where the majority of different classes of objects occupy in the plot.}
    \label{fig:schematic}
\end{figure}

\begin{figure}
	\includegraphics[width=\columnwidth]{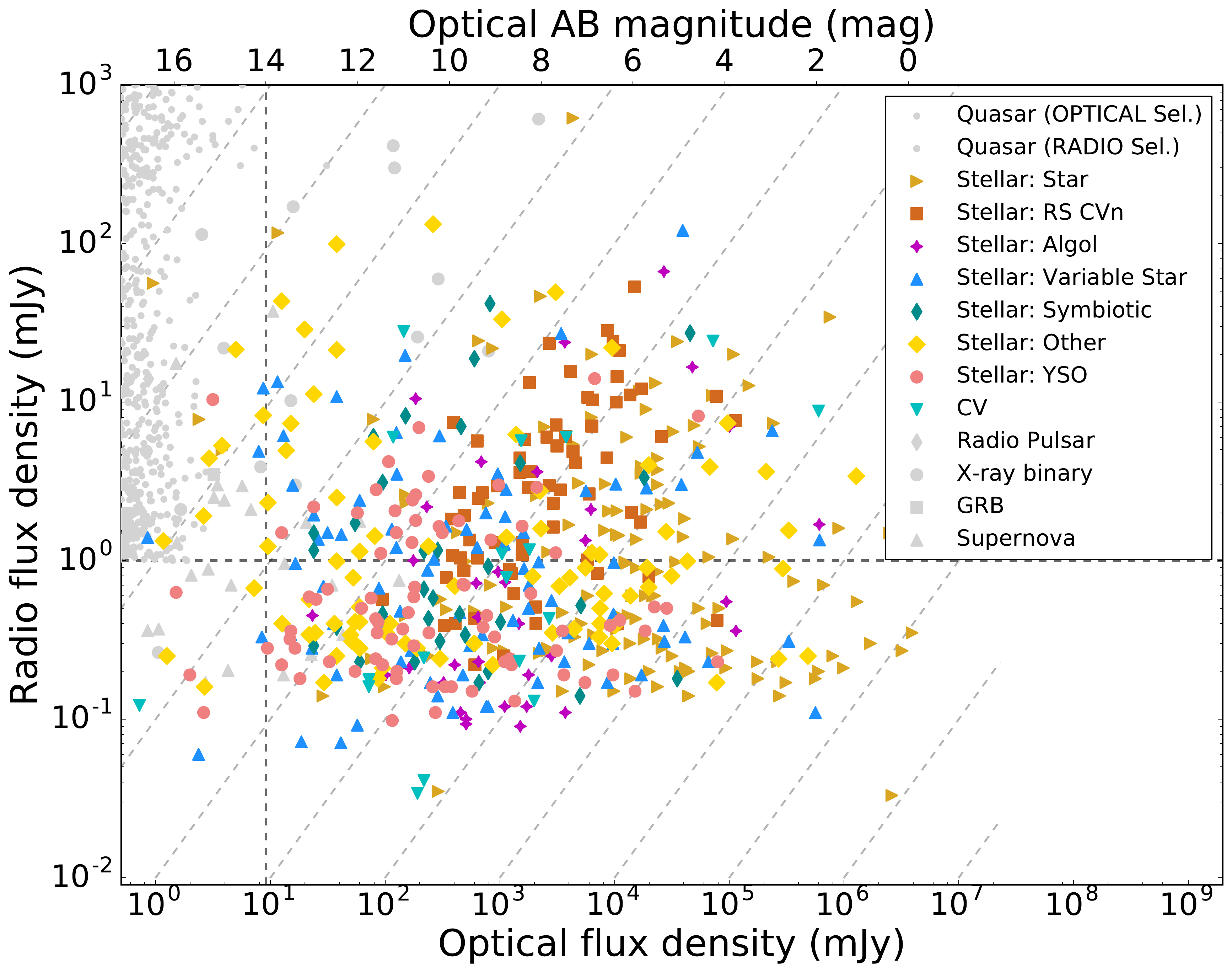}
    \caption{The stellar sample plotted sub-categorised along with CVs on a `greyed out' version of the main plot. It shows that there is no clear trend to be able to use the stellar sub-classifications.}
    \label{fig:stellarsplit}
\end{figure}

\section{Results}
\label{sec:results}
We attempted various methods to visualise the dataset detailed in Section~\ref{sec:method} to find a scheme which offered the best separation between classes. This included using parameters such as the ratio of flux density values, i.e. $F_{\mathrm{r}}/F_{\mathrm{o}}$, and optical colours. We found that plotting $F_{\mathrm{o}}$ against $F_{\mathrm{r}}$, in log space, is the most useful diagram in terms of being able to distinguish the different classes. Further reasoning behind this choice is discussed in Section~\ref{sec:discussion}. The result is presented in Fig.~\ref{fig:basis}.

\subsection{Description of the base result}
\label{sec:describe}
The plot shown in Fig.~\ref{fig:basis} provides an overview of where each class of object, given our data and its limitations, typically lies in the $F_{\mathrm{r}}$-$F_{\mathrm{o}}$ parameter space. We have converted the optical magnitudes to millijansky units in order to make a direct comparison to the $F_{\mathrm{r}}$ measurements (an optical magnitude axis is provided at the top of the figure for reference). This was done using the zero flux point of 3631~Jy as we had converted all magnitudes, where appropriate, to the AB system.

A series of lines are also displayed on the plot. First, there are background diagonal dashed-lines that represent the slope of a constant $F_{\mathrm{r}}$ and $F_{\mathrm{o}}$ ratio (the ratio values are also labelled on the plot for each line). This means that if a particular source was moved closer to Earth, or further away, it would move on the plot along this gradient. Second, the arrow in the lower right-hand section of the plot demonstrates the movement an object could undergo with a reddening equal to 5~mag. Third, and most importantly, there are lines signifying the flux density limits of the surveys used to gather the optically selected quasars, which has a large influence on the result. The horizontal line indicates the 1~mJy source detection threshold of the FIRST survey. The two vertical lines denote the SDSS saturation limit (14~mag, $R$~band) and the SDSS 95 per cent completeness limit (22~mag, $R$~band). As can been seen in Fig.~\ref{fig:basis}, the limits greatly impact the ability to separate classes, especially in the case of the quasars where the abrupt 1~mJy limit of the FIRST survey artificially isolates the GRBs. The reliance on the FIRST survey makes this unavoidable and we specifically address this problem in Section~\ref{sec:distances}.

Considering the limits, Fig.~\ref{fig:split} signifies the region of the plot that is considered the most complete area of the sample, i.e. the area contained by the limits of FIRST and SDSS which are shown by solid red lines, along with sources in the `complete area' being presented in colour. Also on Fig.~\ref{fig:split} we have shown estimates of the region that will be explored by MeerKAT and MeerLICHT, assuming a 1~h, 5$\upsigma$ MeerKAT sensitivity of $23.5$~$\upmu$Jy \citep{MeerKATsensitivity} and a 5~min optical depth of 22.3~mag ($r$-band) for MeerLICHT \citep{meerlichtlimit}. Fig.~\ref{fig:schematic} presents a schematic version of the full area of the plot in Fig.~\ref{fig:basis}, to further illustrate the areas of certain classes. It is created by placing ellipses centred on the median points of the respective populations data and then manually sized to enclose the majority of the populations. As such, Fig.~\ref{fig:schematic} is primarily meant as a visual demonstration, and is presented with the same caviet as discussed above in that the quasars would extend into the GRBs with a more complete sample.
 
From Figs.~\ref{fig:basis}--\ref{fig:schematic}, we see that there is a good separation between Galactic and extragalactic populations in our dataset both in the `complete' and overall regions. On the right half of the plot reside the stellar and CV populations, while classes such as quasars and GRBs typically reside on the left hand side. SNe are mostly placed between these two Galactic and extragalactic populations. The same is true of the XRBs however they are much more chaotic and span a large region that reaches the extragalactic area, highlighting the very dynamic nature of these objects. The Galactic/extragalactic separation is likely due to the detected radio stellar objects in our sample being relatively nearby (hence their detection), therefore they appear bright in the optical. Note that we do not attempt to further sub-classify the stellar sources, using the information from the dataset (see Section~\ref{sec:stellar}). As shown in Fig.~\ref{fig:stellarsplit}, the subclasses of the stellar sample did not offer any distinct separation, hence we remained with the global `stellar' definition. Fainter in $F_{\mathrm{o}}$ are the GRBs which are currently well separated from the rest of the classes due to the FIRST 1~mJy limit as previously discussed (we attempt to address this in Section~\ref{sec:distances}), although the GRBs themselves are relatively clustered together. As for the quasars, there is no distinction between the optically and radio selected samples, with the radio sample occupying the radio-bright section of the former. Hence, from here on we group the two samples under one class of `quasars'. Lastly, the majority of the radio pulsar population are positioned to the far left-hand side of the plot due to their faint $F_{\mathrm{o}}$. The faintness of the radio pulsars in $F_{\mathrm{o}}$ means that they are not contaminated by other classes in our current dataset, although again, this is largely due to SDSS optical limit. We note that the discussed distinctions are primarily due to the $F_{\mathrm{o}}$, though this was our original aim. The $F_{\mathrm{r}}$ becomes important once we consider sources beyond our current limits (Section~\ref{sec:distances}).

\subsection{Dynamic characteristics}
\label{sec:dynamic}
\begin{figure*}
	\includegraphics[width=0.99\textwidth]{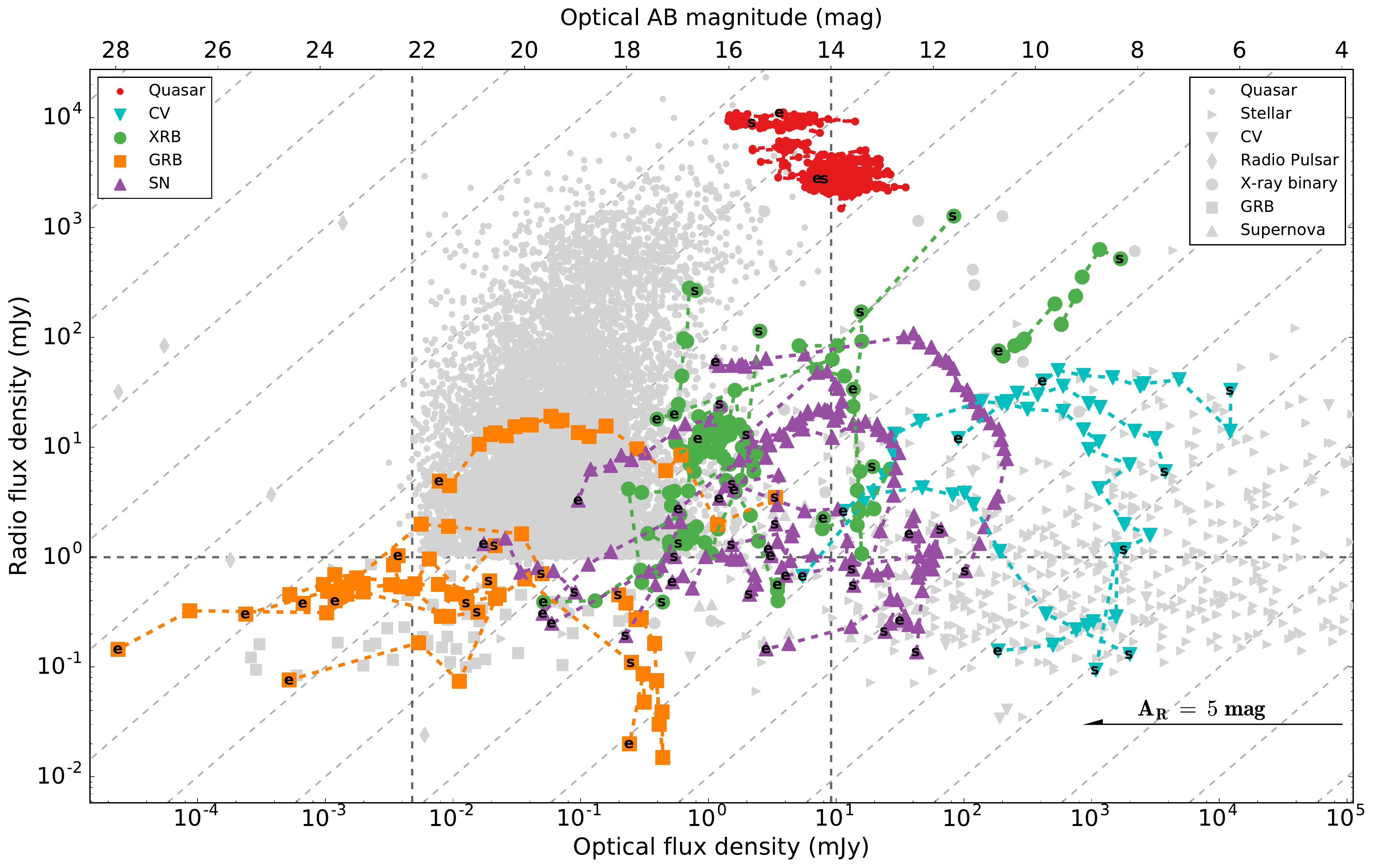}
\includegraphics[width=0.99\textwidth]{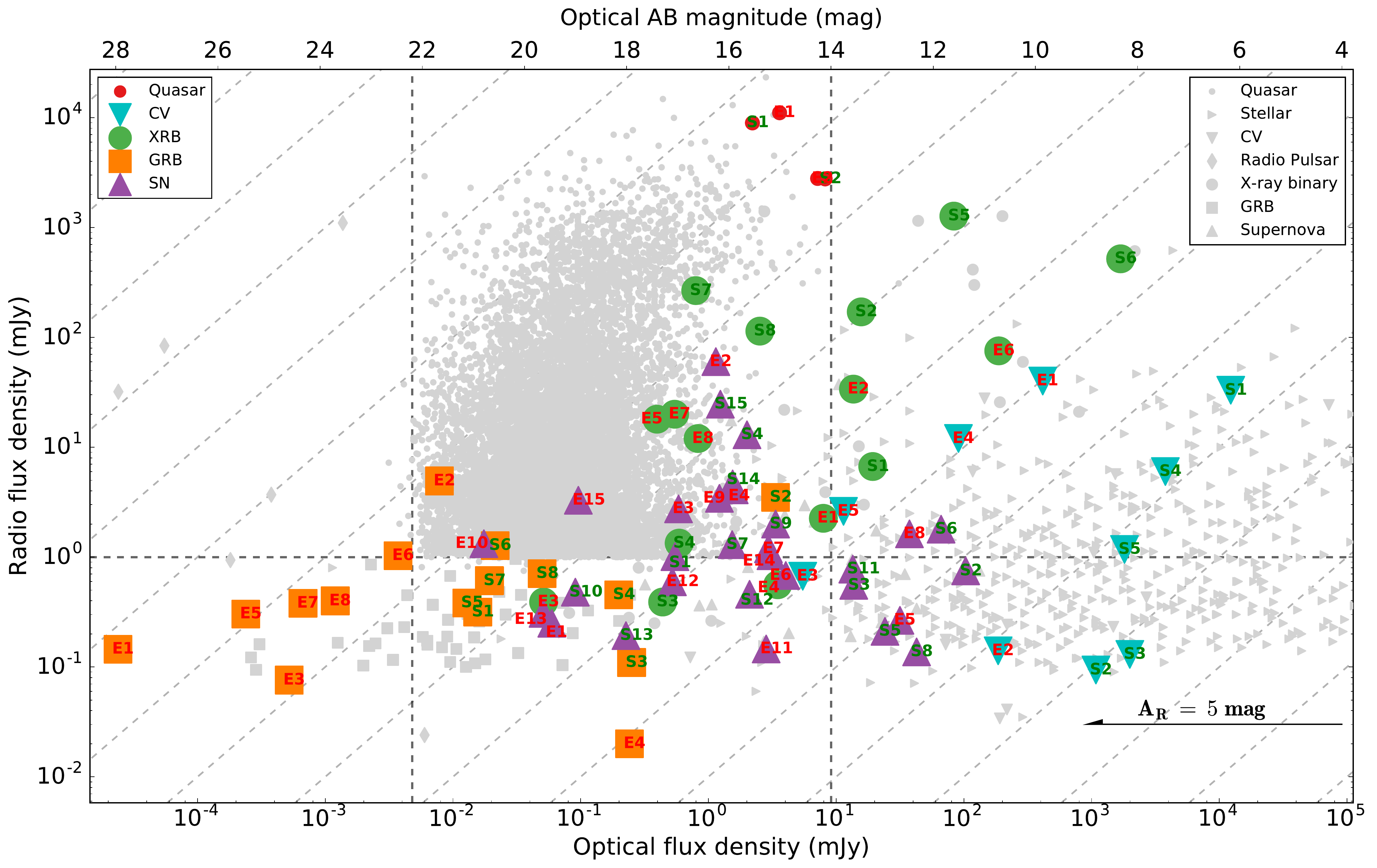}
    \caption{\textit{Upper panel}: Plotted are the `tracks' for which the dynamic sources follow as the astrophysical events evolve over time. Due to the number of sources they are plotted in groups of the respective class of which there are two quasars, five CVs, eight XRBs, 15 SNe and eight GRBs (refer to Table~\ref{table:dynamic} for details of the objects). These are plotted over a `greyed-out' version of the basis diagram to allow comparison. The start and end of each track are labelled with `s' and `e', respectively. \textit{Lower panel}: In this figure only the start and end points of the tracks are plotted for clarity, with each point labelled with a green `S' or red `E'  for the start and end point, respectively, with a corresponding number to pair them. Please refer to the online print for a colour version of the figure.}
    \label{fig:tracks}
\end{figure*}

The results of the dynamic data have been split over two figures: the upper panel of Fig.~\ref{fig:tracks} shows paths which are traced out by the 38 objects, grouped by class, of the dynamic data sample (detailed in Section~\ref{sec:dynamicdata} and Table~\ref{table:dynamic}), overlaid on a `greyed-out' version of Fig.~\ref{fig:basis}. The lower panel of Fig.~\ref{fig:tracks} shows only the `start' and `end' points of the tracks with respective labels linking the data points of an individual source. The nature of the sample means that the `start' and `end' refer to different points in the time line of the event respective to one another. Hence, for a more detailed view, the individual tracks from each class of object (with the exception of the quasars) are shown in Fig.~\ref{fig:individualtracks}. In this figure, the tracks are labelled with the number of days that have elapsed since the first observation in the case of CVs and XRBs, or the number of days after the explosion date for SNe and GRBs (the non-standardised reporting of `day~0' for XRBs and CVs is the reason for the difference). A reminder that the timespan of the objects range from $\sim$2~weeks to 2~yr (with the exception of BL~Lacerate which covers 10~yr) and that each track uses only a single frequency or optical filter. Overall, from Fig.~\ref{fig:tracks} the majority of classes appear to evolve within the static areas as discussed in the previous section (Fig.~\ref{fig:basis} and \ref{fig:schematic}). This can largely be seen by looking at the `start' and `end' points in the lower panel of Fig~\ref{fig:tracks}. The exception to this is the SNe which brighten in the radio considerably during their evolution. The XRBs also still have a large overlap with other classes, especially SNe. Below we expand on the features of each class.
\subsubsection{Accreting sources}
\label{sec:accretingdynamic}
Starting with the two quasars, for which 3C~454.3 is the source brighter in the radio with BL~Lacertae below, it can be seen that these move very little over time in $F_{\mathrm{r}}$ relative to the other classes presented in the figure. In both cases, the two quasar objects do not show large relative flux density variation events like those seen in other classes, in addition to varying back and forth rather than following a clear track. 3C~454.3 is monitored for 630~d which is similar order of magnitude to the timespan covered by three CVs and one SN, which show much more variation over the same timescale. As the quasars are quite static on the plot they are not displayed in Fig.~\ref{fig:individualtracks}.

The CVs appear quite uniform in their dynamic characteristics. Initially detected as bright optical objects, the brightest in the dynamic sample, they proceed to decrease in $F_{\mathrm{o}}$ while the $F_{\mathrm{r}}$ rises and falls again. The sources follow a simple model in which the optical flare evolves more rapidly than the radio resulting in an `anti-clockwise loop' appearance of the track. This is a characteristic that is shared with other classes on the plot such as GRBs and SN. Hence, anti-clockwise loops are a common occurrence. The `start' of the measurements in the sample occurs 1--2~months after the optical detection, i.e. when the first radio detections were made, and therefore the optical is seen to decline. The exception to this is SS~Cyg, the dwarf nova, that was observed in radio rapidly ($1.3$~d) after the detection of an optical flare. The radio flare from SS~Cyg also rose and faded quickly as seen by its track.

The tracks of the XRBs are less uniform and reflect the overall sample characteristic of appearing at a diverse range of areas on the plot, both in start and end points. As stated in Section~\ref{sec:dynamicdata}, most of the sources are observed in a declining state after the detection of an X-ray flare, and a rapid decline in both $F_{\mathrm{r}}$ and $F_{\mathrm{o}}$ is what is most commonly seen on the plot. Although the $F_{\mathrm{o}}$ of GS~1124-684 remains steady during the radio decay and experiences a re-flaring in the radio. A rise in $F_{\mathrm{r}}$ is detected in three sources: GRO~J1655-40, GX~339-4 and XTE~1859+226. GX~339-4 after the radio rise mimics the steady decline of the other sources, however the other two sources show more complex behaviour. GRO~J1655-40 continues to rise in $F_{\mathrm{o}}$ while the $F_{\mathrm{r}}$, rises, peaks and decays all during this period, hence why it is the only XRB to end brighter in $F_{\mathrm{o}}$ in the sample. XTE~1859+226 on the other hand experiences many smaller radio flares during the optical decay, hence its erratic appearance, though it should be noted that this is data from the Green Bank Interferometer (GBI) database and therefore the sampling of the radio light-curve is very frequent compared to other sources (the source contains 42 pairs of measurements over 2~months compared to 16 for the next most sampled source).

\subsubsection{Explosive events}
\label{sec:accretingdynamic}
SN would be expected to follow the anti-clockwise motion as discussed with the CVs, and two SN show a very good example of that, 1993J (type II) and 1994I (Ic). They both rise steadily in the radio band, with only a slight increase,
\begin{landscape}
\begin{figure}
	\includegraphics[scale=0.31]{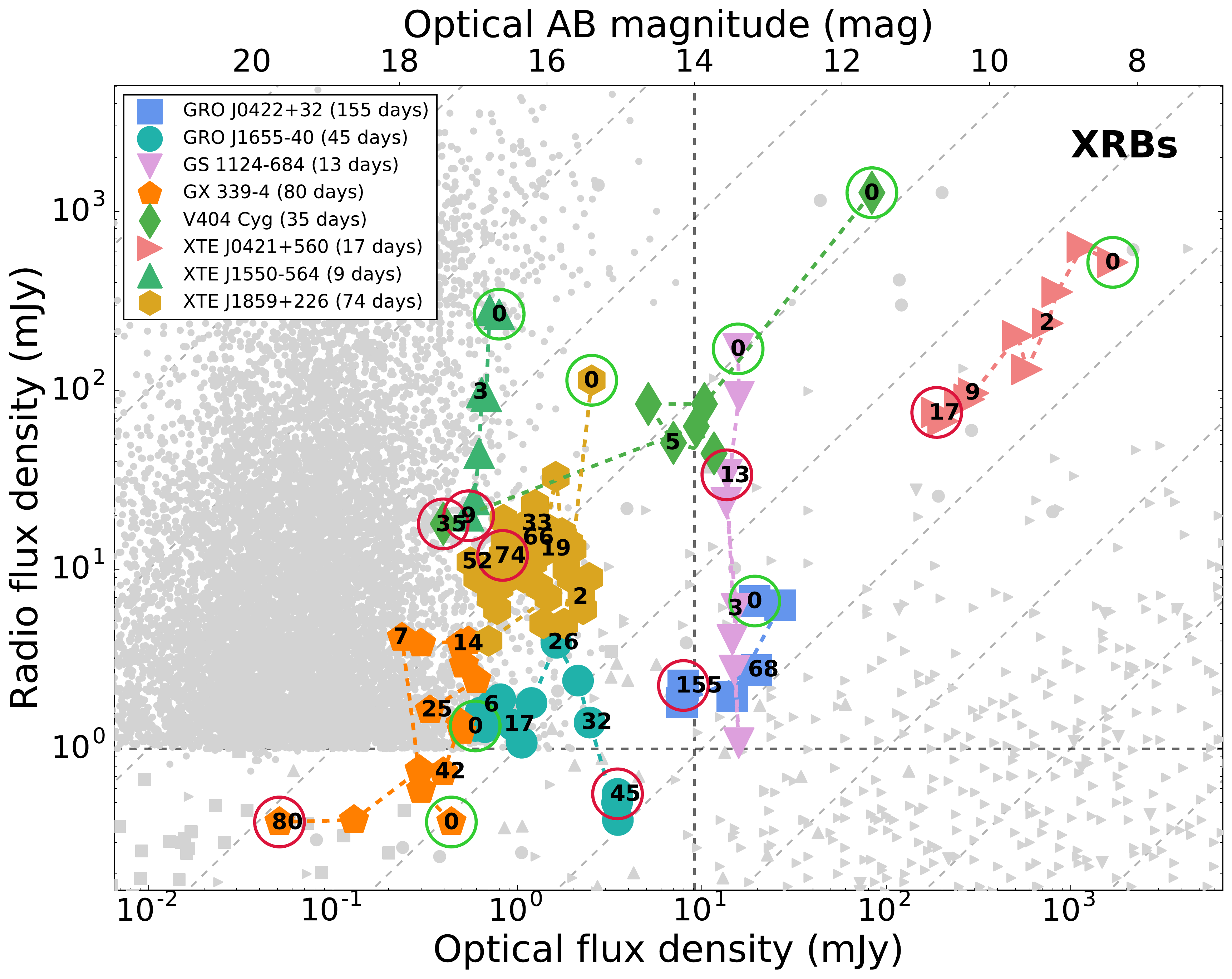}
\vspace{0.3cm}
\includegraphics[scale=0.31]{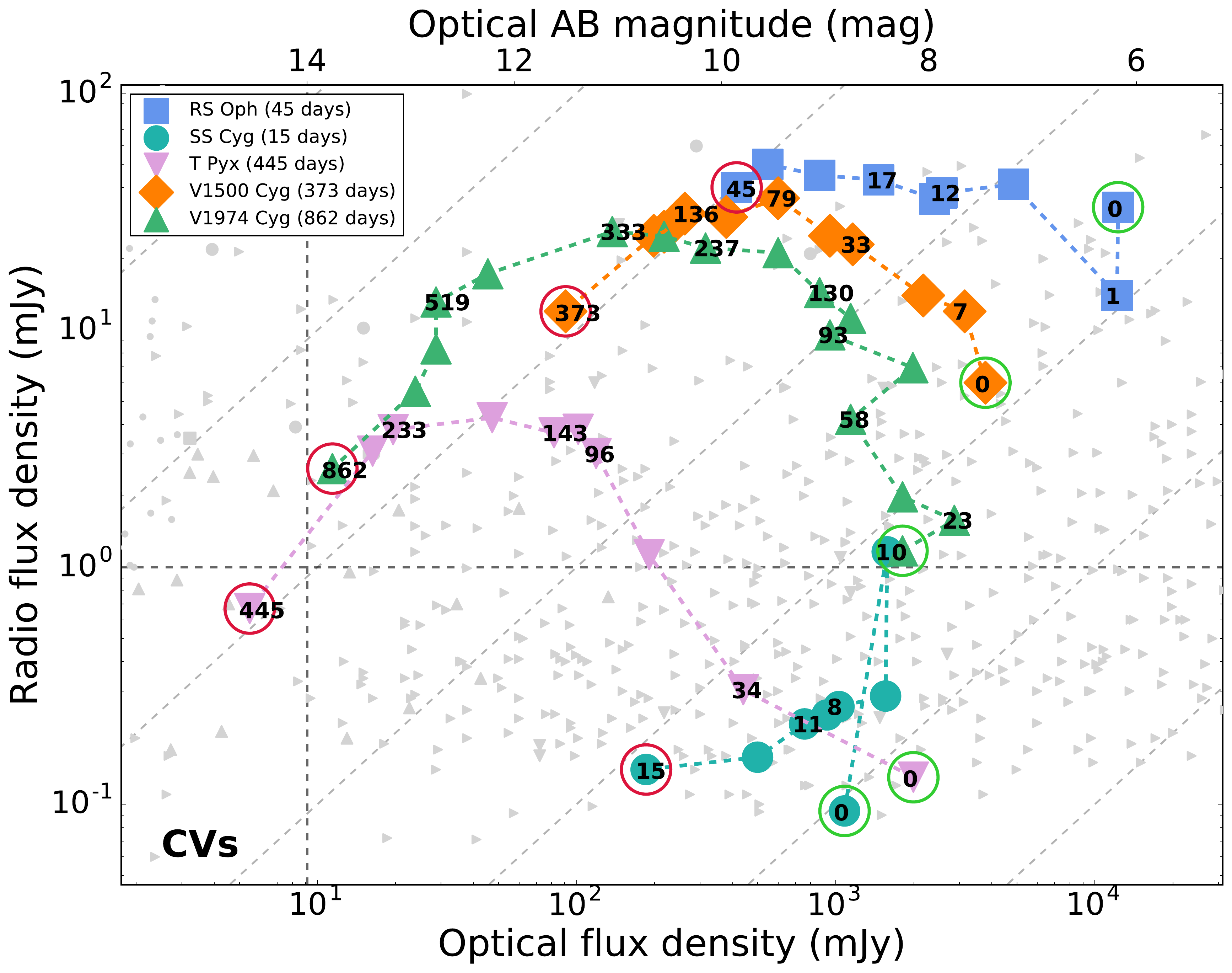}
	\includegraphics[scale=0.31]{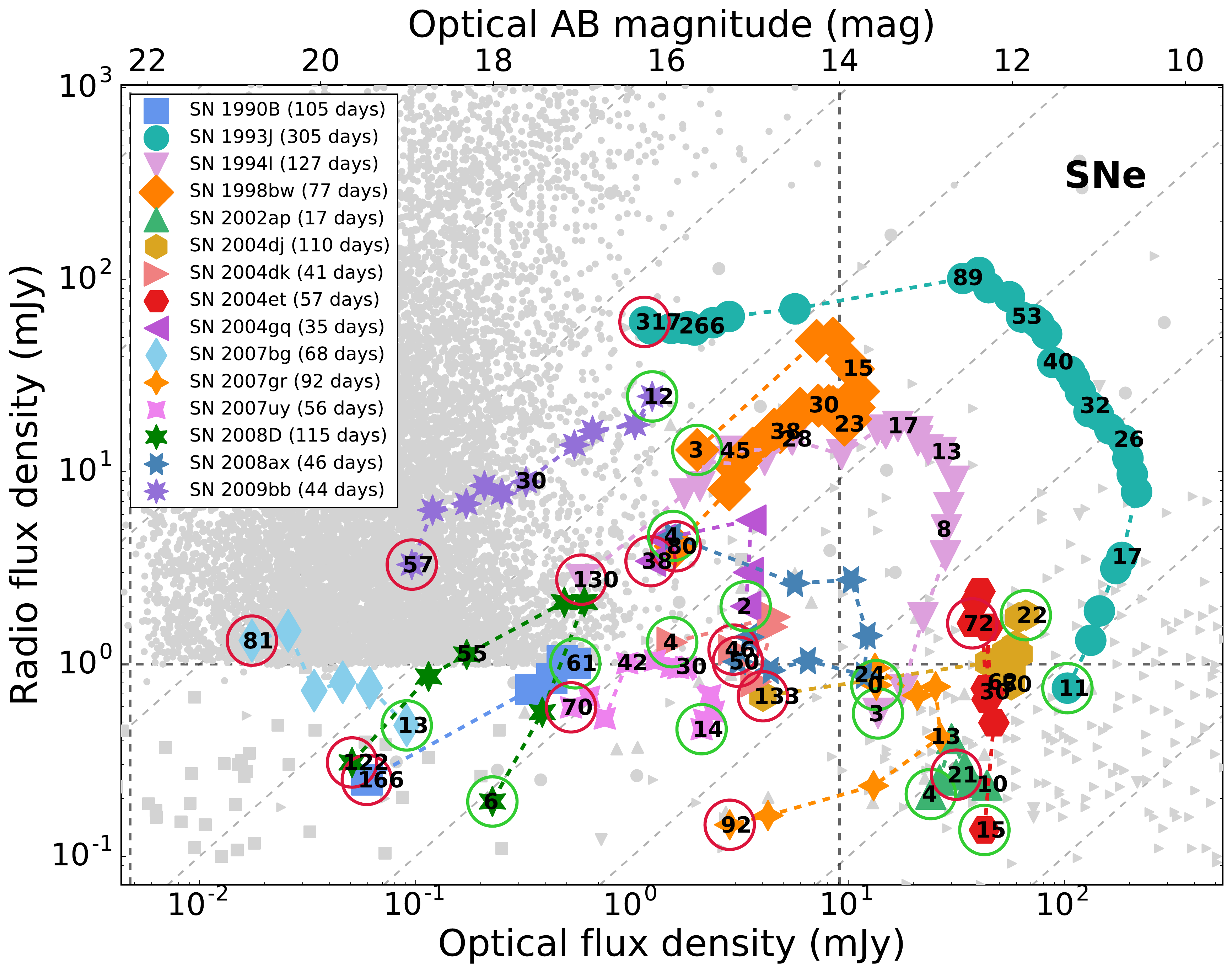}
\hspace{1.2cm}
	\includegraphics[scale=0.31]{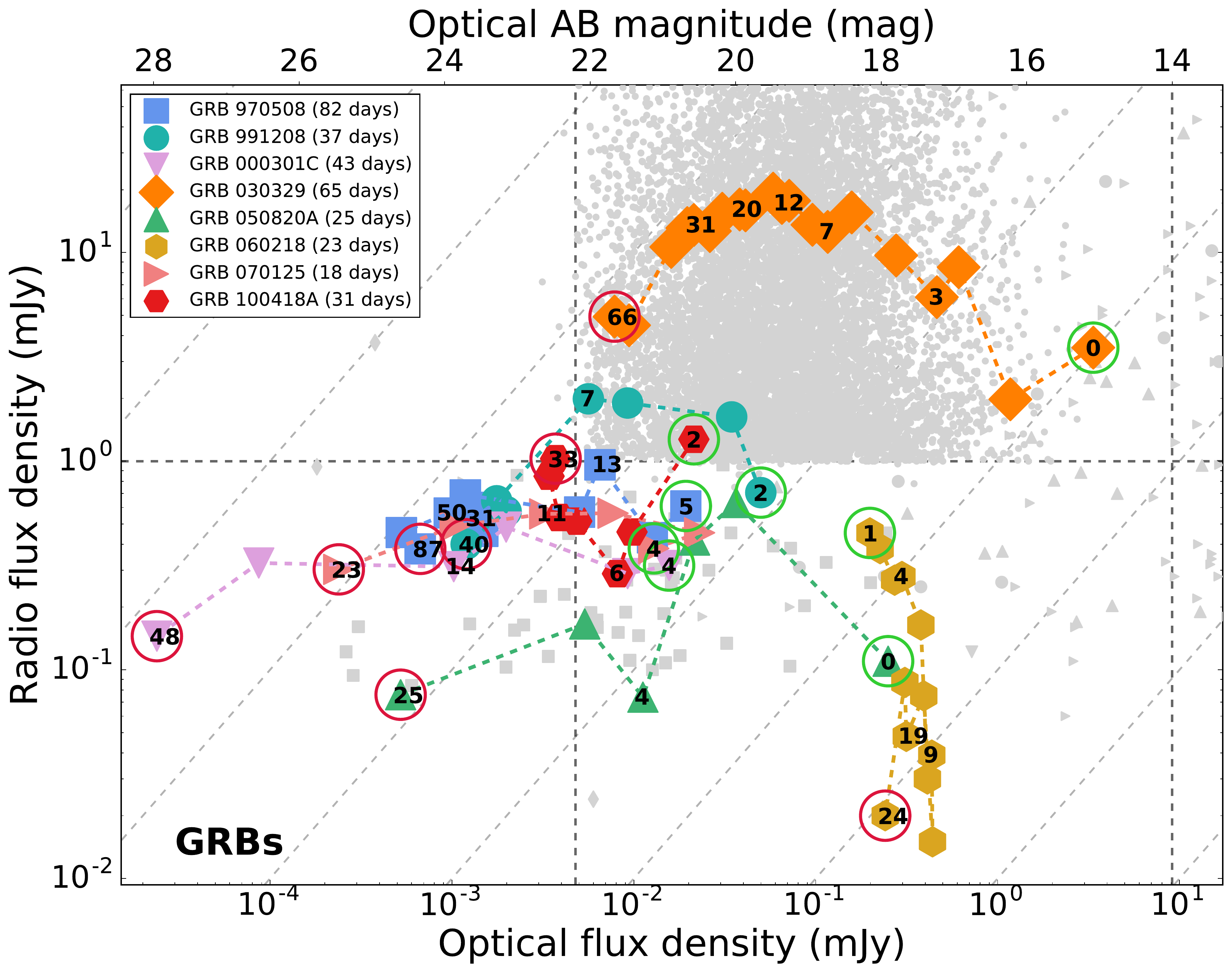}{\,\,}
    \caption{The tracks of individual sources in the dynamic sample plotted per class. The start and end of each track is denoted by a green and red ring around the data point, respectively. Also on every start, end and every third data point (this frequency varies due to avoid congestion on the plot) is the number of days that has passed. For CVs and XRBs the day 0 refers to the first observations, however for GRBs and SNe it refers to the explosion date. \textit{Upper left}: XRBs, \textit{upper right}: CVs, \textit{lower left}: SNe and \textit{lower right}: GRBs. Please refer to the online print for a colour version of the figure.}
    \label{fig:individualtracks}
\end{figure}
\end{landscape}
\noindent and then decrease, in the optical band. The reverse scenario then becomes true for the later part of the tracks where the $F_{\mathrm{r}}$ changes slower than the now decreasing $F_{\mathrm{o}}$. Generally, SNe follow this trend, and from the first pair of measurements, the track will tend to proceed in the upper-right direction, i.e. brighter in $F_{\mathrm{r}}$ and $F_{\mathrm{o}}$ before looping around in the anti-clockwise motion. However, there are some SNe that behave differently. For example, SN~1998bw (Ic) moves instead in a tight clockwise motion. This is due to SN~1998bw experiencing a rapid rise to its peak $F_{\mathrm{r}}$ ($\sim$10~d), compared to, for example, SN~1994I ($\sim$20~d) as seen in Fig.~\ref{fig:individualtracks}. SN~2009bb, another type Ic, also shows a steady decline after 12 days, suggesting a similar event to that of SN~1998bw. SN~2007bg (Ic) is unique in that it appears very near the GRB region and also initially mimics the track `shape' of other GRB sources. A year after discovery, persistent strong radio emission detected by \citet{SN2007bg_offaxis} suggested that SN~2007bg could be a good candidate for an off-axis GRB. Although this was later deemed unlikely with further analysis by \citet{SN2007bg_r}, the plot could prompt this suggestion from the track shape and placement.

The GRBs also follow an anti-clockwise motion, though with a much more muted rise in $F_{\mathrm{r}}$ than seen in SNe, and also are mainly all contained within the previously defined GRB area of the plot. Two sources stand out: GRB~030329, which appears to be very bright in the radio, and GRB~060218 that does not follow the common GRB track trajectory of a slowly decaying radio source. Instead it decays in the radio very swiftly. First, GRB~030329 was a nearby afterglow \citep[$z=0.1685$, 587~Mpc;][]{030329_redshift} and at the time was the brightest GRB afterglow in the radio ever observed. Thus, its position on the plot is boosted in $F_{\mathrm{r}}$ such that it transits the quasar area. If the GRB was moved further away, e.g. the same as GRB~100418A \citep[$z=0.624$, 1.4~Gpc;][]{100418A_redshift}, the 10~mJy flux density becomes $\sim$1~mJy - consistent with the other GRBs in the sample. Due to its brightness and well sampled light-curve, the anti-clockwise nature of the track can be clearly seen. GRB~060218 on the other hand is defined as an X-ray flash (XRF), and had unique properties in that it was relatively nearby ($z=0.0335$, $\sim$145~Mpc) and the burst itself lasted a long time ($\upDelta t\approx$~2,000~s). Indeed, \citet{GRB060218_r} showed that the event was a hundred times less energetic but ten times more common than cosmological GRBs.

The dynamic data sample has shown how taking into account how a discovered object evolves over time - considering the time scale of change and the positioning - could potentially aid the ability to separate classes where confusion occurs. This is discussed further in the following sections.

\section{Discussion}
\label{sec:discussion}
We have presented a representation of the $F_{\mathrm{r}}$ and $F_{\mathrm{o}}$ measurements for a wide range of transient and variable sources, showing that it is possible that this information alone can differentiate between Galactic and extragalactic populations. Here we give more detailed discussion of the obtained results.

First, given how the data is presented in Fig.~\ref{fig:basis}, we believe the range of radio frequencies does not have a negative impact. For example, assuming a canonical spectral index of $\alpha=-0.8$ \citep[for optically thin synchrotron emitting sources;][]{radiospectralindex}, a 1~Jy source at 1~GHz becomes $\sim$0.2~Jy at 10~GHz, a one order of magnitude change. A difference of two orders of magnitude would mean $\alpha=-2.0$. Hence, given the populations we are sampling (likely $\alpha\sim-0.8$), a spread of one order of magnitude does not impact our ability to differentiate classes. We also note that we have not considered measurement uncertainties when compiling our results. We believe that because of the nature of what we are investigating, the intrinsic spread of each population in the $F_{\mathrm{r}}$-$F_{\mathrm{o}}$ parameter space has a much greater influence than that of measurement error. We also acknowledge that the dataset created in this investigation is a biased sample, due to the many different sources used to build it. At this initial stage of the project, we do not attempt to un-bias the data due to the complexity of the task. Though, we acknowledge that correcting for the biases is an important future step in constructing more detailed models from the dataset. Lastly, we have not corrected our optical sample for extinction. If this was accounted for, it would cause sources to shift left along the x-axis only, as denoted by the red arrow on Fig.~\ref{fig:basis}. We did not wish to `de-redden' our sample as we wanted to probe measurements that are directly observed, such that the dataset could potentially be used as soon as observations are recorded.
\begin{figure}
	\includegraphics[width=\columnwidth]{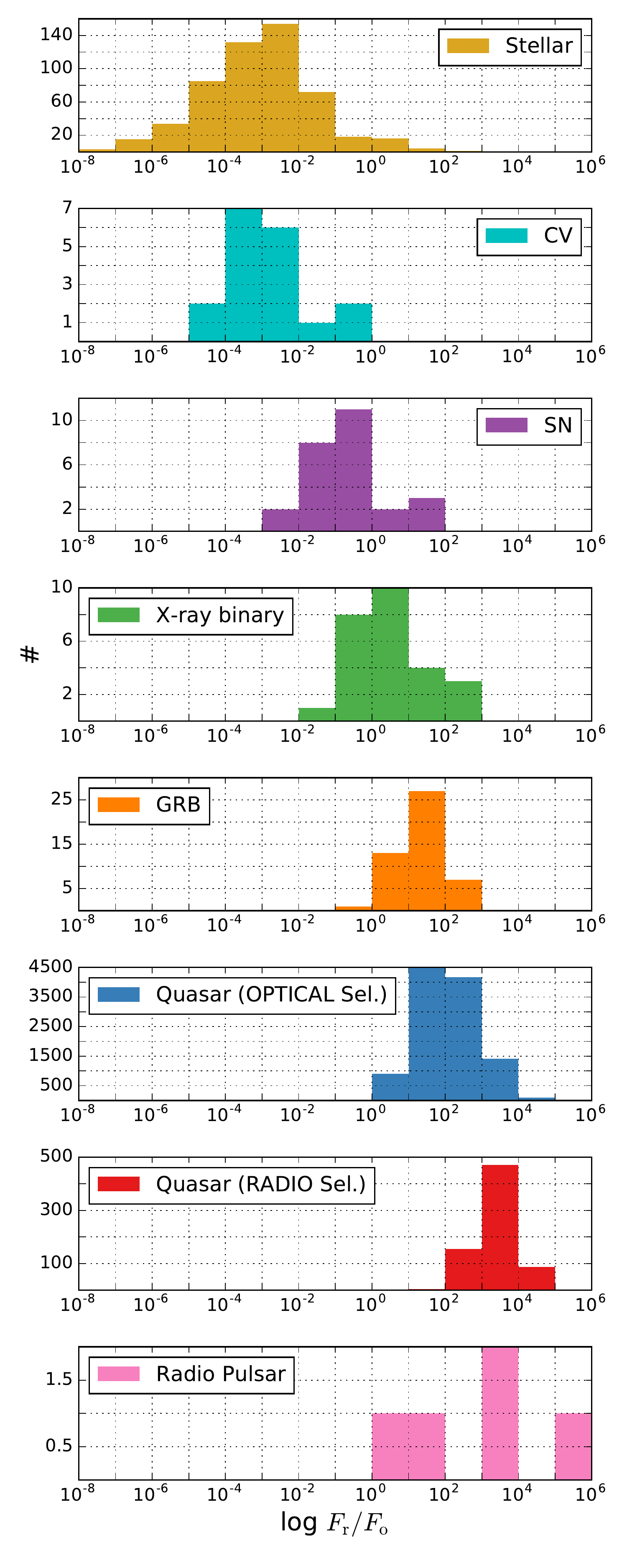}
    \caption{The distribution of the ratio $F_{\mathrm{r}}/F_{\mathrm{o}}$ for each class included in the sample.}
    \label{fig:histogram}
\end{figure}
\begin{figure*}
	\includegraphics[width=\columnwidth]{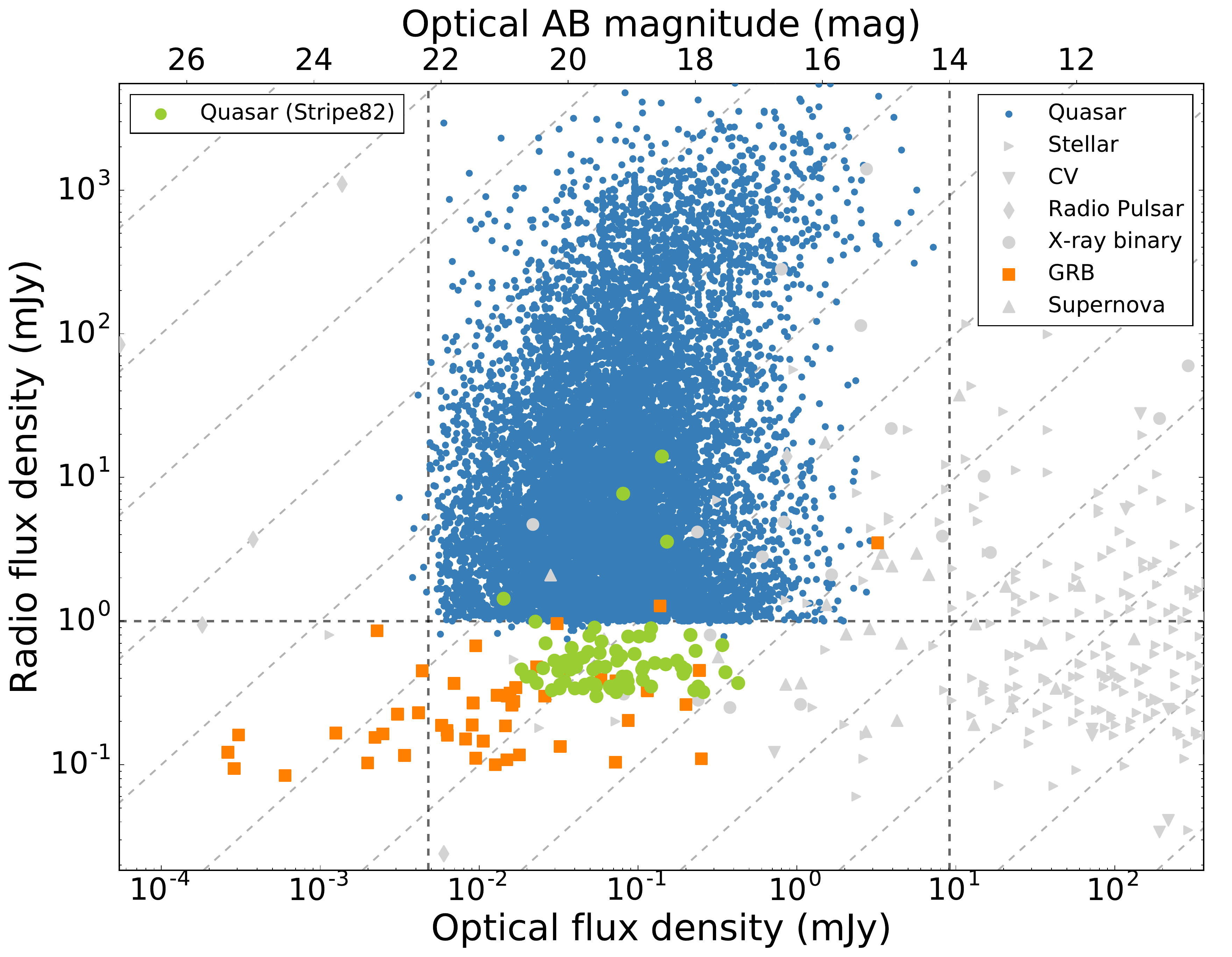}
	\includegraphics[width=\columnwidth]{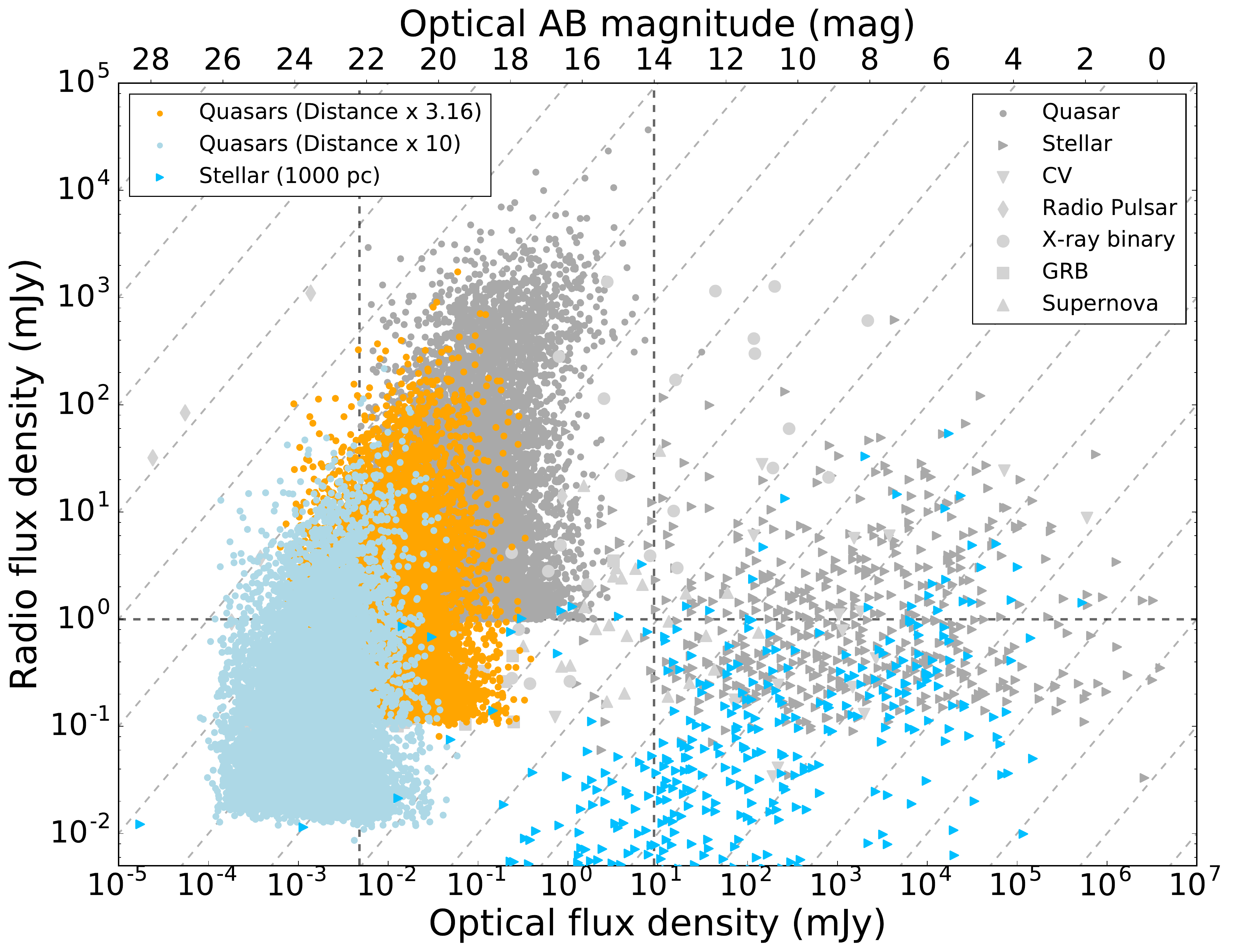}
    \caption{\textit{Left}: The faint quasars of the VLA Stripe~82 survey have been added to the plot where the other sources have been `greyed out' with the exception of the sample quasars and the GRBs. It shows how the fainter quasars overlap with the previously isolated GRBs. \textit{Right}: The effect of probing the quasar and stellar populations to a deeper sensitivity, by increasing the distance of those sources contained in the sample. This is overplotted on a `greyed-out' version of the basis plot with the original quasar and stellar sources highlighted by a darker shade. Please refer to the online print for a colour version of the figure.}
    \label{fig:distances}
\end{figure*}

As for the different parameters considered discussed earlier in Section~\ref{sec:results} - we first considered using the ratio of $F_{\mathrm{r}}/F_{\mathrm{o}}$ and plotting this against $F_{\mathrm{r}}$. As can be interpreted from our final choice, this ratio value also allowed us to somewhat separate classes, especially in the case of Galactic or extragalactic objects. Fig.~\ref{fig:histogram} shows the distribution of $F_{\mathrm{r}}/F_{\mathrm{o}}$ for each class of object. However, this representation did not offer any advantages over plotting the $F_{\mathrm{r}}$ and $F_{\mathrm{o}}$ against each other and hence was not used in the final plot. Another parameter that was tested was optical colour, which we had originally attempted to obtain, although the total percentage of the sample for which colour was obtained varied largely per class (99~per~cent of the quasar sample, 75~per~cent of the SNe, 40~per~cent of CVs, 25~per~cent of the GRBs and 20~per~cent of the XRBs). We found that plotting optical colour versus $F_{\mathrm{r}}$, or versus the ratio of $F_{r}/F_{o}$, greatly hindered our ability to separate different classes of objects. This was primarily caused by the narrow range of optical colour values, which limited the parameter space in which the classes could be separated. For example, with this parameter we could no longer distinguish Galactic and extragalactic populations. Hence, due to this and the limited available data we did not pursue optical colour any further at this time.

\subsection{Extending the sample beyond the current limitations}
\label{sec:distances}
As discussed previously, the location of quasars on the plot is currently not a complete representation and greatly impacts the ability to identify areas where classes are separated. This is due to the sensitivity limits of the surveys we have used to build our sample, in particular, the SDSS 95 per cent completeness limit of $\sim22$~mag, and the FIRST source detection threshold of 1~mJy (refer to Section~\ref{sec:quasars}). The populations would exceed these limits when observed with improved sensitivities, and therefore extend into areas where other classes are present - causing confusion. A similar notion could be applied to the stellar sample, as this will be another general class of a large size that will continue to be detected as observations probe deeper.

Extending the quasar sample with telescope data is currently challenging given the lack of wide-field radio surveys that probe deeper than the FIRST limit. One of the larger surveys that has probed deeper is a survey of the SDSS `Stripe 82' area of sky at high-resolution using the VLA at 1.4~GHz done by \citet{Stripe82VLA}. The survey covers 92~deg$^2$ with a median rms noise of 52~$\upmu$Jy~beam$^{-1}$. The SDSS~Stripe~82 survey consists of 275~deg$^2$ of sky that was visited multiple times by the SDSS camera, and hence reaches $\sim2$~mag deeper than the areas with single visits \citep{SDSSStripe82}. As part of the radio survey, \citet{Stripe82VLA} discovered 76 new radio quasars by cross matching with the SDSS data. The left panel of Fig.~\ref{fig:distances} shows the discovered faint quasars added to our plot, where it can be seen that the sources appear directly below our quasar sample, where previously the GRBs were isolated.

While the Stripe~82 quasars begin to show how the quasars will extend, we can further demonstrate the effect of deeper sensitivities by showing how the current sample would appear if the distance from Earth was increased. The optically selected SDSS quasars are shifted such that the flux density measurements are approximately 10 and 100 times fainter (moved in distance by a factor of $\sqrt{10}$ and $10$ respectively). As the redshifts for the quasars are known from SDSS, we were also able to apply a K-correction and correct for bandwidth stretching. The luminosity with the corrections applied is
\begin{equation}
L=4 \pi F d_\textrm{L}^{2} (1+z)^{-\alpha-1}\textrm{,}
\end{equation}
where $L$ is the luminosity, $F$ is the flux density, $d_\textrm{L}$ is the luminosity distance, $z$ is the redshift and $\alpha$ is the power-law spectrum index. For the radio we use the previously defined canonical value of $\alpha_\textrm{r}=-0.8$ and the for the optical a value of $\alpha_\textrm{o}=-0.44$ \citep{quasaropticalalpha}.
For the stellar population, our sample includes parallax distance measurements for 451 of the 534 stellar objects plotted, for which the median value is 142~pc. Thus, we use this information to increase the distance of these stellar objects by 1~kpc. The results of these shifts can be seen in the right panel of Fig.~\ref{fig:distances}.

The extension of the quasars completely covers the area of the plot that was previously only occupied by GRBs, along with approaching the pulsar region. This suggests that when observing beyond the limits of our current sample, it will prove difficult to differentiate a GRB from a quasar with a single pair of measurements. The stellar sources extend down to an $F_{\mathrm{r}}$ of 1~$\upmu$Jy, with some sources going beyond this value. The region below the stellar sources on the basis plot was empty, hence, unlike the quasars, the extension of the stellar sources does not cause confusion with other sources. It is worth noting that the distinction between quasars and stellar sources, i.e. Galactic and extragalactic objects, remains relatively clear. However, the $F_{\mathrm{r}}$ becomes much more relevant when determining the Galactic--extragalactic boundary. The extended stellar sources have similar $F_{\mathrm{o}}$ to that of the original quasar sample, hence the $F_{\mathrm{r}}$ becomes a useful differentiator between the classes. 

\subsection{Transient diagnostic application}
\label{sec:classifier}
\begin{figure*}
	\includegraphics[width=\textwidth]{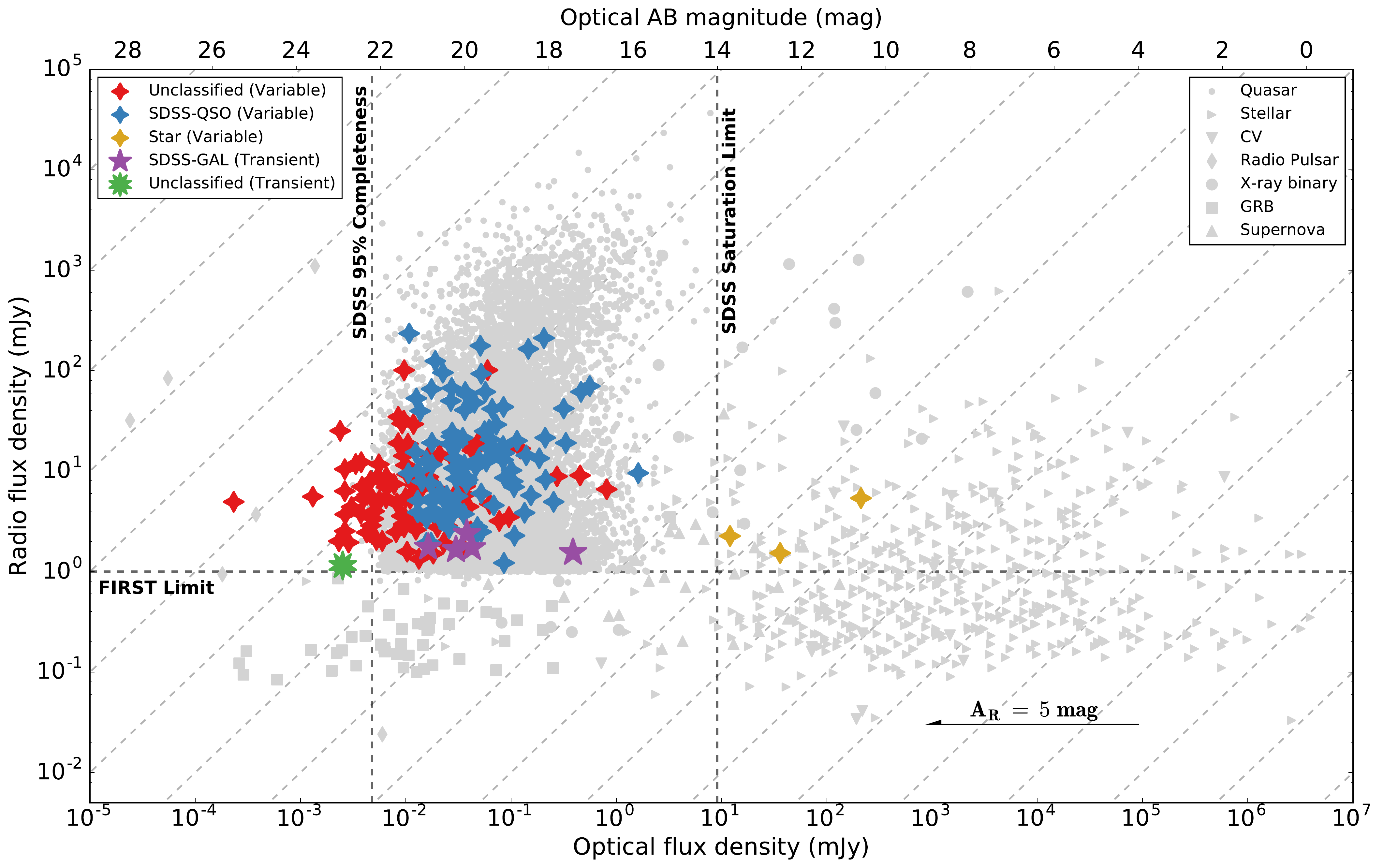}
    \caption{The result of plotting the selected radio transients and variables from the FIRST survey \citep{FIRSTvariables}, categorised by their highest ranked counterpart at other wavelengths deduced by the authors of the catalogue, upon a `greyed-out' version of the basis plot. Please refer to the online print for a colour version of the figure.}
    \label{fig:test}
\end{figure*}

To help demonstrate the use of the current dataset to provide an initial class approximation of radio sources, we perform a small scale blind test using a catalogue of variable and transient radio sources discovered in the FIRST survey, as compiled by \citet{FIRSTvariables} (hereafter the `T11 sample'). We impose the following criteria on the sources we select from the catalogue: 
\begin{enumerate}
\item sources must have a defined SDSS $i$-band magnitude, which are mostly gathered by the authors from the SDSS DR7 catalogue. SDSS-$r$ band would be more ideal but for the purposes of the test, plus to keep within the already done cross matching, the SDSS-$i$ is adequate enough for our use.
\item only sources which are not flagged as being in the vicinity of another source brighter than 500~mJy.
\end{enumerate}
The authors also attempted to classify the candidates by cross-matching with a variety of catalogues and databases. The sources are labelled as either transient or variable. For the test we select the candidates that have the classifications as defined by the authors of:
\begin{description}
\item \textbf{SDSS-*} - an unclassified SDSS counterpart (transient and variable).
\item \textbf{SDSS-QPO (S) and (P)} - quasars from the SDSS spectroscopic \citep[S;][]{FIRST_SDSS_S} and photometric \citep[P;][]{FIRST_SDSS_P} quasar catalogues (only contain variables).
\item \textbf{SDSS-GAL} - objects classified as galaxies in SDSS (select transients only).
\item \textbf{Tycho-Star} - counterpart found in the Tycho star catalogue \citep[][only contain variables]{tycho}.
\item \textbf{HIP-Star} - counterpart found in the Hipparcos star catalogue \citep[][only contain variables]{hip}.
\end{description}
We do not consider SDSS-Gal variables due to these sources effectively being already classified, in addition to our previously stated desire to focus on optically point-like transient or variable objects. However, it should be noted that radio transients may appear coincident with extended optical objects, but can be difficult to associate the two unambiguously. Hence, this scenario would not be ignored in a blind search. For the purposes of our test we define our own categories of:
\begin{description}
\item \textbf{Unclassified (Variable)} (SDSS-* variables).
\item \textbf{SDSS-QSO (Variable)} (SDSS-QPO S,P variables).
\item \textbf{Star (Variable)} (Tycho/HIP-Stars, variables).
\item \textbf{SDSS-Gal (Transient)}.
\item \textbf{Unclassified (Transient)} (SDSS-* transients).
\end{description}
In total we include 198 objects in our testing sample, the by-category totals can be found in Table~\ref{table:test}. We plot the $F_{\mathrm{r}}$ and $F_{\mathrm{o}}$ values of the 198 transient and variable objects onto the basis diagram, for which the result is shown in Fig.~\ref{fig:test}.

\begin{table}
	\centering
	\caption{The total number of each type of object included in the test sample.}
	\label{table:test}
	\begin{tabular}{cc} 
		\hline
		Object Type & Total No.\\
		\hline
		Unclassified (Variable) & 89\\
		SDSS-QSO (Variable) & 100\\
		Star (Variable) & 3\\
		SDSS-Gal (Transient) &5\\
		Unclassified (Transient) & 1\\
		\hline
		Total & 198\\
		\hline
	\end{tabular}
\end{table}
We see an agreement with the classified `Star' and `QSO' objects, in that they lie within the stellar and quasar regions respectively, a distinction that could have been possible without the prior classifications. This is the immediate first advantage of the basic method - the separation between Galactic and extragalactic objects - at least in the sample we currently have available. It is a promising first step of using such a photometric method to provide a swift classification, and also a useful property. In a scenario with a high yield of transient sources, the basic distinction could be used to direct further follow-up. Although, classifying deeper than this is challenging, and in the current form of the method, attributing a definitive class is not possible. Without the prior classifications from the T11 sample, it would be difficult to firmly identify an object that might not belong to the quasar population. Though, it should be noted that the majority of these sources are likely to be quasars, and are classified as such by both methods. One exception is the optically very faint `Unclassified variable' that is noticeably separate from the main quasar cloud near the pulsar region, but even so, it is again more than likely to be part of a optical faint quasar population. Given the current data, this would be an example of a source to be initially followed up. If the T11 sample was not using the FIRST survey then we may have seen some objects below 1~mJy, making such objects more likely to belong to the GRB population. Unlike the sub-stellar populations, it may be that the extragalactic areas of the plot have areas of a specific class, such as GRBs or SN, which do seem to be grouped, despite the inevitable extension of the quasar population (Section~\ref{sec:distances}).

\subsection{Future steps}
\label{sec:future}
Building a probability model using the dataset is the natural progression of this work, and would continue to see advancement and refinement with the addition of further data over time. Accounting for the bias of the extremely diverse dataset is, again, important for this step, but complex. Though, in its current form, we believe the dataset already offers usefulness in gaining an initial diagnostic.

Combining this method with other techniques and external knowledge can only enhance its accuracy. We previously showed how a sample of sources evolve on the plot over time (Section~\ref{sec:dynamic}). A more thorough analysis of this feature may unveil trends unique to certain classes that could be recognised by an algorithm within the $F_{\mathrm{r}}$ and $F_{\mathrm{o}}$ parameter space, meaning that new sources could be classified with a high probability and minimal monitoring time (though the `minimal monitoring time' would also need to be established). This would also open up cross examining the rise/decline times of the events in question with existing classification techniques both the radio \citep{Pietka,PietkaClassifier} and optical \citep{opticaltime}. Knowledge such as whether the optical counterpart is an extended object, or if the object was present in historical catalogues is also simple information that offer clues as to the associated class. In time, the addition of a third parameter (e.g. X-ray flux density) could also be achieved.

We stress and remind that we are assuming an era where the rate of detected radio transients and variables is much greater than it is at present. Obtaining a spectrum of a counterpart will usually provide the most definitive clue of an objects associated class. Though with potentially hundreds or thousands of candidates, a spectra may not be feasible to obtain swiftly in order to be fed into ongoing observation strategies with specific science goals. The presented $F_{\mathrm{r}}$--$F_{\mathrm{o}}$ method is purposely simple for it is an initial analysis of what can be achieved with current $F_{\mathrm{r}}$--$F_{\mathrm{o}}$ data. The simple test performed above shows that it is difficult to obtain precise classifications outside of being Galactic or extragalactic. But, advancing the $F_{\mathrm{r}}$--$F_{\mathrm{o}}$ dataset itself, along with combining the method with other information and techniques, can provide an important initial diagnostic method in a future classification system.

\section{Conclusions}
\label{sec:conclusions}
We have presented an initial study of the optical counterparts to a wide range of radio transients and variables, with the eventual aim of using such a dataset to help classify a blindly detected radio event. We collected a sample consisting of 12,441 pairs of $F_{\mathrm{r}}$ and $F_{\mathrm{o}}$ measurements from seven classes of astrophysical objects, including populations such as quasars, GRBs, CVs and SNe. These measurements were recorded by numerous radio and optical facilities, at frequencies ranging from 1--10~GHz and optical filters $V$ and $R$. The dataset has a significant bias due to the diverse range of classes of object and sources of information. However, our approach at this stage was to provide an early assessment, meaning that we do not yet tackle the complex biases.

We found that comparing the $F_{\mathrm{r}}$ and $F_{\mathrm{o}}$ measurements of each object provided a useful separation between Galactic and extragalactic sources. Radio emitting stars and CVs, for example, are distinctly in a different area to that of quasars and GRBs. However, differentiating between distinct individual classes with a high accuracy is challenging. Types of stellar sources were found to be impossible to separate, but pulsars and extragalactic sources such as GRBs were found to be isolated. Although, the isolation is caused by the limitations of our sample, specifically the detection limit of the FIRST survey used for quasars. Fainter radio quasars beyond this limit will lead to confusion with previously isolated classes. We also investigated how a sample of 5 transient/flare CV events, 8 XRB transient/flare events, 15 SNe, 8 GRBs and 2 quasars evolved in the parameter space over time - a possible unique feature that could be used to provide more accurate classifications with minimal monitoring. We showed that CVs, SNe and GRBs follow similar `tracks' in the parameter space, in that they generally follow an `anti-clockwise' loop on the plot - a motion that would be repeated for any class that follows a model of the optical flare evolving more rapidly than the radio. XRBs on the other hand are quite varied in nature where as quasars tend to move relatively little over the same time frame. Lastly we used a sample of radio transients and variables identified in the FIRST survey to highlight the features discussed above.

The next step of our investigation will be to use our dataset to provide a probability of an object belonging to a class, for which it is also important to understand the biases. Though in its current form, the dataset can already provide a useful first estimate of a discovered source, albeit crude. However, once used alongside other classification techniques and information it is likely to help provide accurate predictions. In addition, the method can also be iterated on and improved as further simultaneous multi-wavelength data is recorded. For example, the MeerLICHT project, which will provide simultaneous optical data to MeerKAT radio observations, is a unique opportunity for a detailed dataset in the near future. A need for a classification system assumes an era in the future where a high number of radio transients and variables are regularly being blindly detected. Ideally these systems would be automatic, likely using machine learning techniques and subsequently able to make appropriate decisions on further follow-ups depending on the science goals.

\section*{Acknowledgements}
We thank the anonymous referee for their detailed comments that helped to improve our analysis and presentation.
We acknowledge support from the European Research Council via Advanced Investigator Grant no. 267697 4 Pi sky: Extreme Astrophysics with Revolutionary Radio Telescopes (PI: R.P. Fender). T.M.-D. also acknowledges support via a Ram\'on y Cajal Fellowship (RYC-2015-18148) and the the Spanish MINECO grant AYA2017-83216-P.

The authors wish to thank Manuel~G\"{u}del for providing us with the formatted data on the flux density characteristics of the stellar sources.

This research has made use of data taken and assembled by the WEBT collaboration and stored in the WEBT archive at the Osservatorio Astrofisico di Torino - INAF (\url{http://www.oato.inaf.it/blazars/webt/}). We thank Claudia~Raiteri and Massimo~Villata for supplying the WEBT AGN monitoring data.

The authors also wish to thank Deanne~Coppejans for providing extra information on the CVs used in the sample.

We acknowledge with thanks the variable star observations from the AAVSO International Database contributed by observers worldwide and used in this research.

This paper has made use of up-to-date SMARTS optical/near-infrared light-curves that are available at \url{www.astro.yale.edu/smarts/xrb/home.php}. The Yale SMARTS XRB team is supported by NSF grants 0407063 and 070707 to Charles~Bailyn.

This research has made use of the SIMBAD database and the VizieR catalogue access tool, operated at CDS, Strasbourg, France.

The Green Bank Interferometer is a facility of the National Science foundation operated by NRAO with support from the NASA High Energy Astrophysics program.

Funding for the SDSS and SDSS-II has been provided by the Alfred P. Sloan Foundation, the Participating Institutions, the National Science Foundation, the U.S. Department of Energy, the National Aeronautics and Space Administration, the Japanese Monbukagakusho, the Max Planck Society, and the Higher Education Funding Council for England. The SDSS Web Site is http://www.sdss.org/.

The SDSS is managed by the Astrophysical Research Consortium for the Participating Institutions. The Participating Institutions are the American Museum of Natural History, Astrophysical Institute Potsdam, University of Basel, University of Cambridge, Case Western Reserve University, University of Chicago, Drexel University, Fermilab, the Institute for Advanced Study, the Japan Participation Group, Johns Hopkins University, the Joint Institute for Nuclear Astrophysics, the Kavli Institute for Particle Astrophysics and Cosmology, the Korean Scientist Group, the Chinese Academy of Sciences (LAMOST), Los Alamos National Laboratory, the Max-Planck-Institute for Astronomy (MPIA), the Max-Planck-Institute for Astrophysics (MPA), New Mexico State University, Ohio State University, University of Pittsburgh, University of Portsmouth, Princeton University, the United States Naval Observatory, and the University of Washington.

Funding for SDSS-III has been provided by the Alfred P. Sloan Foundation, the Participating Institutions, the National Science Foundation, and the U.S. Department of Energy Office of Science. The SDSS-III web site is http://www.sdss3.org/.

SDSS-III is managed by the Astrophysical Research Consortium for the Participating Institutions of the SDSS-III Collaboration including the University of Arizona, the Brazilian Participation Group, Brookhaven National Laboratory, Carnegie Mellon University, University of Florida, the French Participation Group, the German Participation Group, Harvard University, the Instituto de Astrofisica de Canarias, the Michigan State/Notre Dame/JINA Participation Group, Johns Hopkins University, Lawrence Berkeley National Laboratory, Max Planck Institute for Astrophysics, Max Planck Institute for Extraterrestrial Physics, New Mexico State University, New York University, Ohio State University, Pennsylvania State University, University of Portsmouth, Princeton University, the Spanish Participation Group, University of Tokyo, University of Utah, Vanderbilt University, University of Virginia, University of Washington, and Yale University.



\bibliographystyle{mnrasedit}
\bibliography{opticalradio} 




\appendix

\section{Sample frequency/band distribution and reference tables}
\label{app:tables}
\begin{figure}
	\includegraphics[width=\columnwidth]{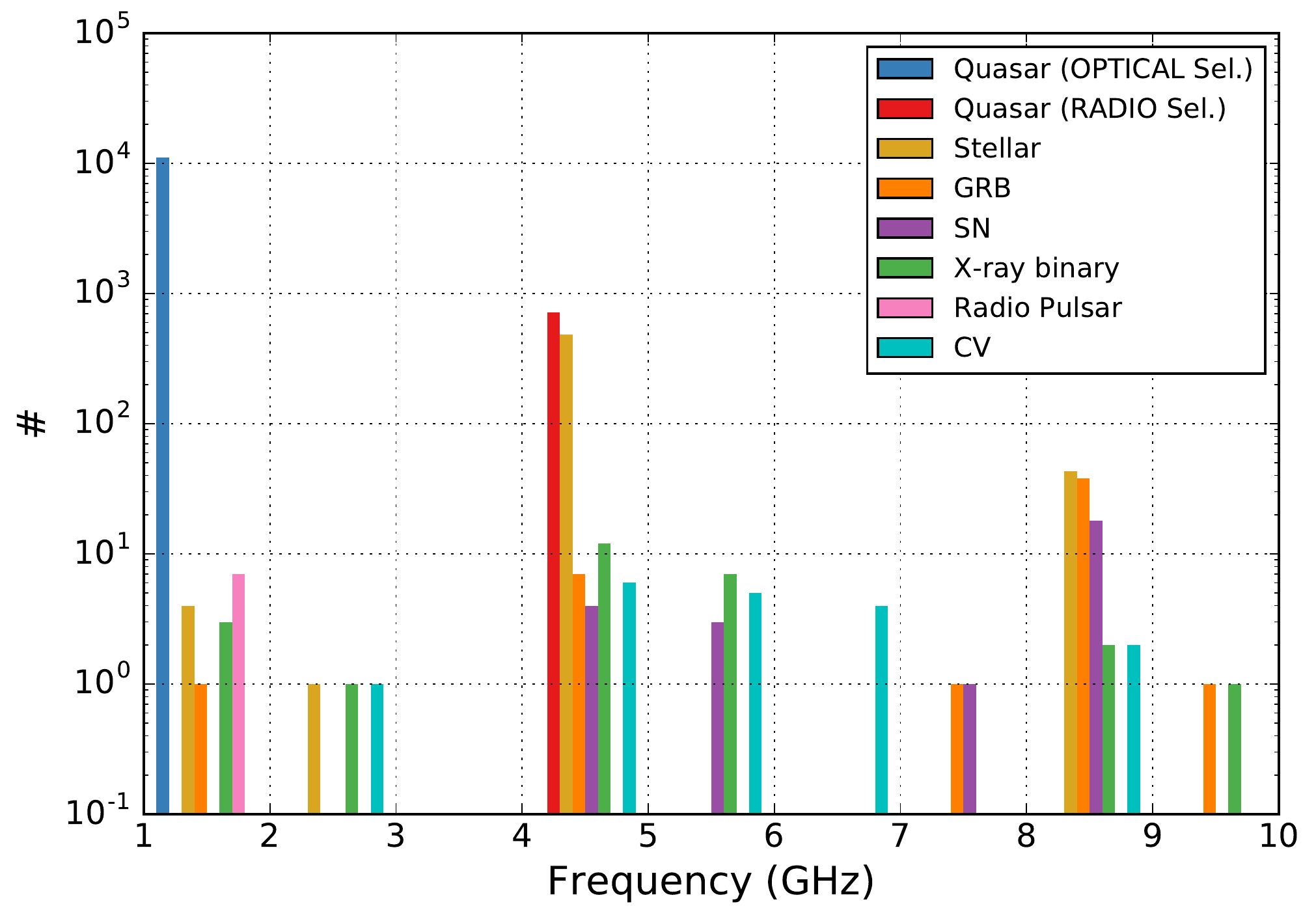}
	\includegraphics[width=\columnwidth]{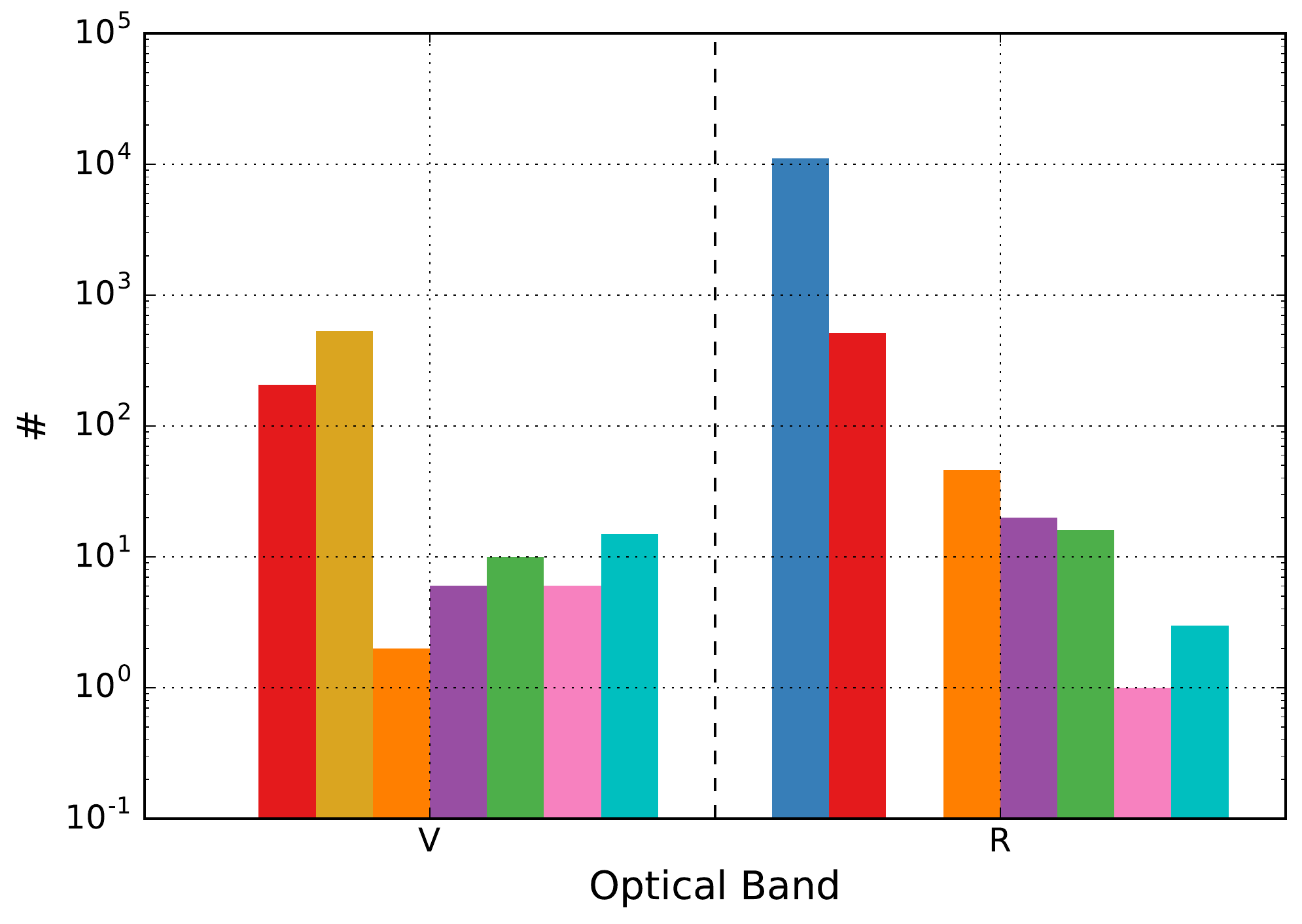}
    \caption{\textit{Upper panel}: The distribution of radio frequencies used for the $F_{\mathrm{r}}$ measurements collected for each class of object in the sample. Each bin is a width of 1~GHz. \textit{Lower panel}: The distribution of optical bands used for the $F_{\mathrm{o}}$ measurements collected for each class of object in the sample. The legend in the upper plot also applies to the lower.}
    \label{fig:histograms}
\end{figure}

\begin{figure}
	\includegraphics[width=\columnwidth]{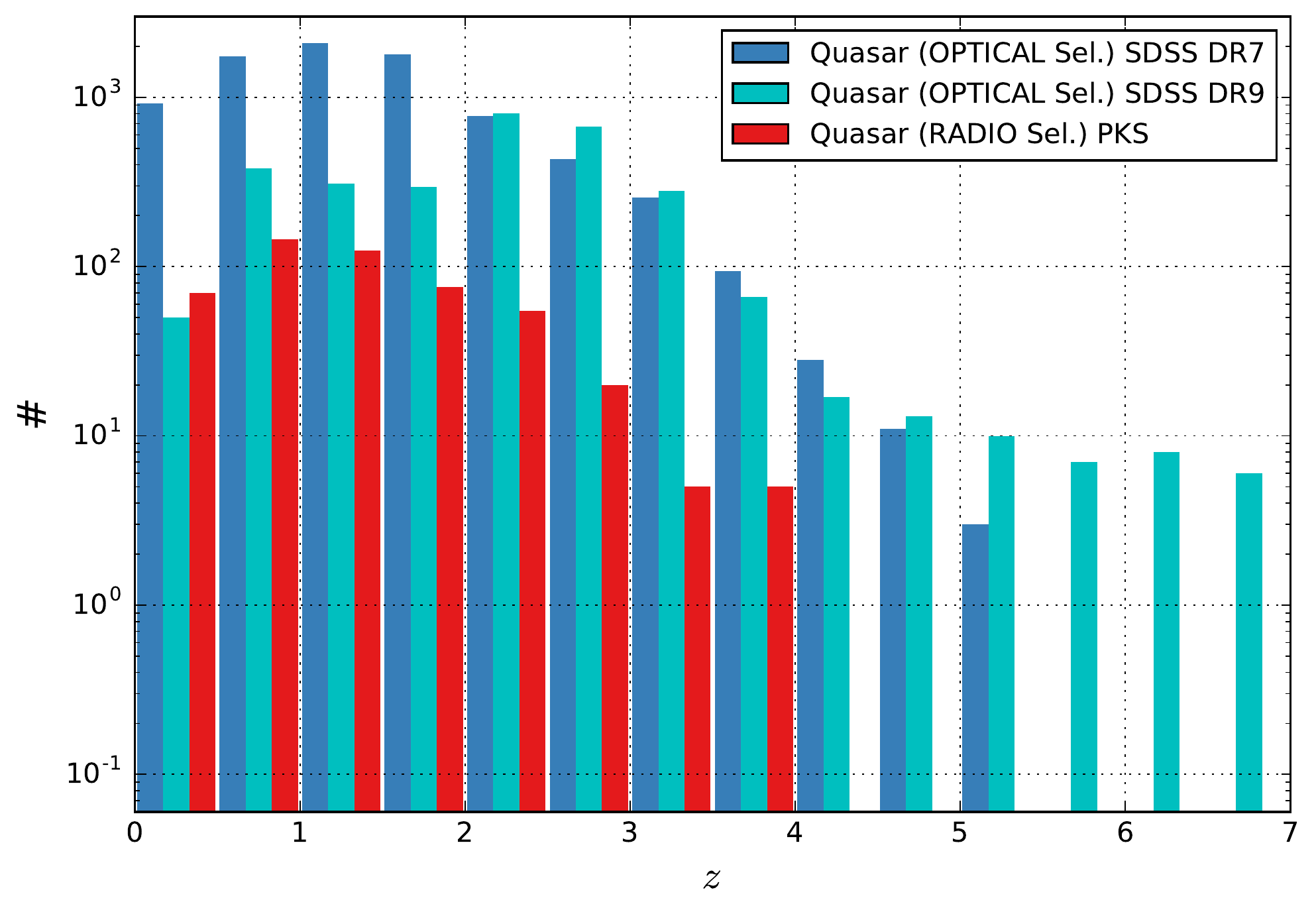}
    \caption{The distribution of redshifts for the quasar sample, complied from the Seventh Data Release quasar catalogue \citep[DR7QC;][]{SDSSQCDR7}, the Ninth Data Release quasar catalogue \citep[DR9QC;][]{SDSSQCDR9} and the Parkes Survey \citep{PKS}. Note that not all redshifts were available for the Parkes survey.}
    \label{fig:qsozhistogram}
\end{figure}

\begin{figure}
	\includegraphics[width=\columnwidth]{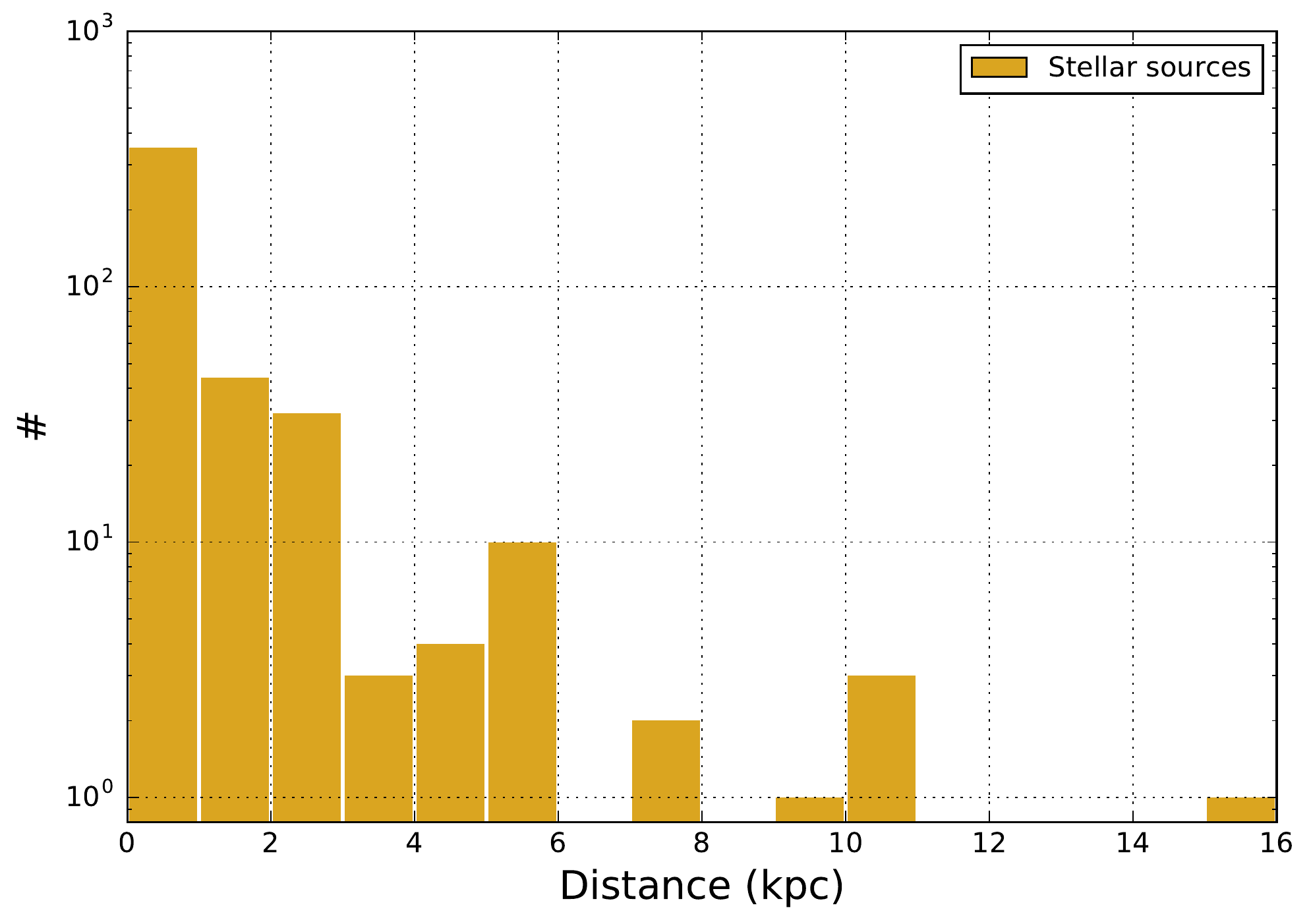}
    \caption{The distribution of the parallax distance measurements of the stellar sample compiled using the data presented in \citep{Gudel}.}
    \label{fig:stellardisthistogram}
\end{figure}

\begin{figure}
	\includegraphics[width=\columnwidth]{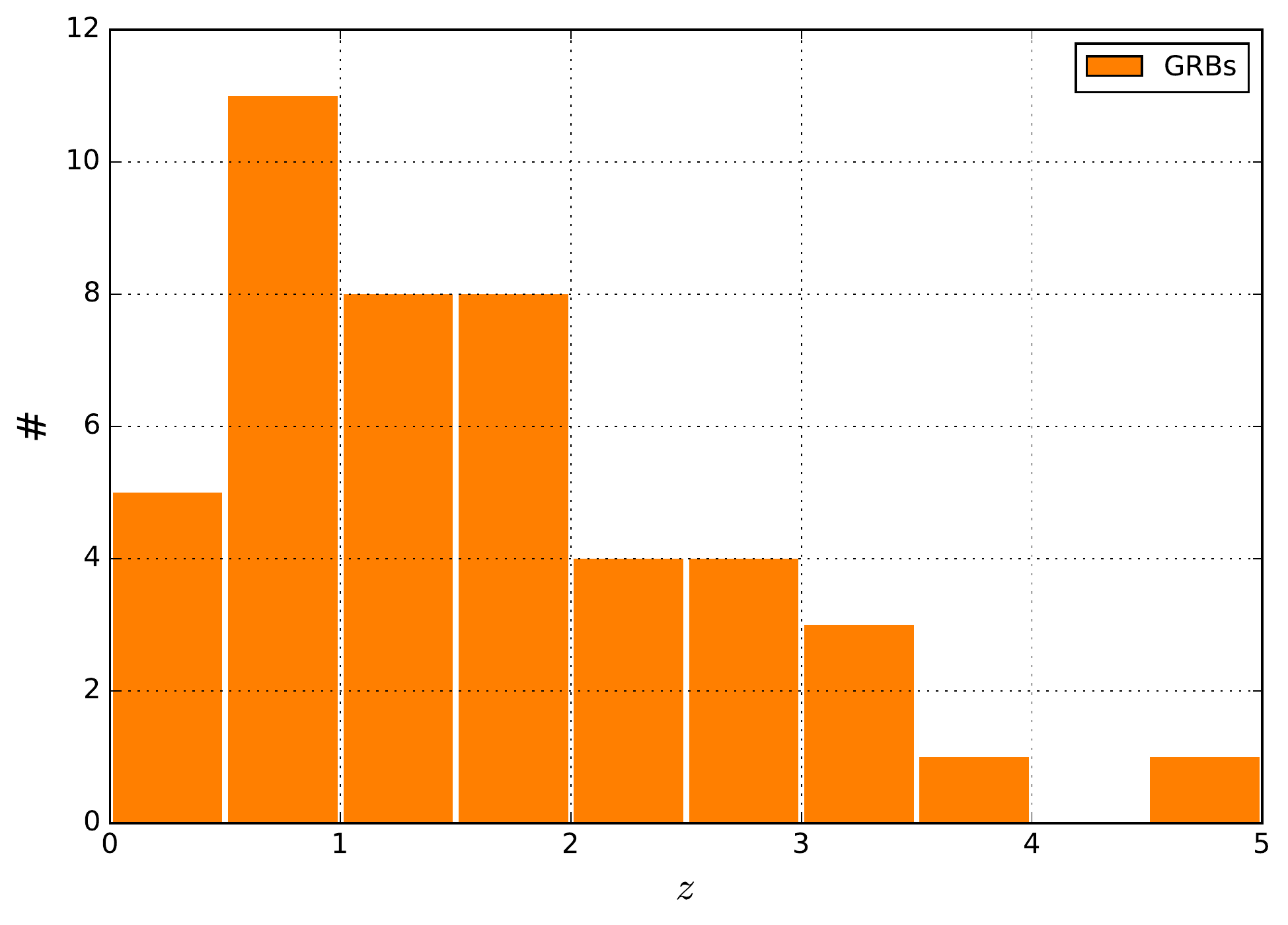}
    \caption{The distribution of redshifts of the GRB sample. Refer to Table~\ref{table:GRBs} for the references of the redshift measurements.}
    \label{fig:grbzhistogram}
\end{figure}

Fig.~\ref{fig:histograms} shows the distribution of the radio frequencies and optical bands used for the sample, for each class of object. The redshift distribution of the quasar sample is shown in Fig.~\ref{fig:qsozhistogram}, followed by the distribution of the parallax distance measurements of the stellar sample in Fig.~\ref{fig:stellardisthistogram}. Fig.~\ref{fig:grbzhistogram} shows the redshift distribution of the GRB sample. Tables~\ref{table:pulsars}, \ref{table:CVs}, \ref{table:x-ray}, \ref{table:SNe} and \ref{table:GRBs} provide the references for the $F_{\mathrm{o}}$ and $F_{\mathrm{r}}$ measurements for our sample of radio pulsars, CVs, XRBs, SNe, and GRBs, respectively.

\begin{table*}
	\centering
	\caption{The radio pulsars that are included in the sample. The radio measurements and distances were gathered with the use of the ATNF Pulsar Catalogue \citep{pulsar_cat}.}
	\label{table:pulsars}
	\begin{tabular}{cccc} 
		\hline
		Object Name & $F_{\mathrm{o}}$ ref. & $F_{\mathrm{r}}$ ref. & d (kpc)\\
		\hline
		Crab (B0531+21) & \citet{pulsars_optical} & \citet{crab_r} & 2.00\\ 
		Vela (B0833-45) &\citet{pulsars_optical} & \citet{vela_r} & 0.28\\
		B0540-69  &\citet{pulsars_optical} & \citet{B0540-69_r} & 49.70\\
		B1509-58  & \citet{pulsars_optical}& \citet{B1509-58_r} & 4.40\\
		B0656+14  &\citet{pulsars_optical} & \citet{crab_r} & 0.29\\
		B1133+16  &\citet{pulsars_optical} & \citet{crab_r} & 0.35\\
		B0950+08  & \citet{pulsars_optical}& \citet{crab_r} & 02.6\\
		\hline
	\end{tabular}
\end{table*}

\begin{table*}
	\centering
	\caption{The CVs that are included in the sample. The first section contains CVs obtained from the Stellar sample (see Section~\ref{sec:stellar}). The second section contains CVs obtained from the data contained in \citet{Pietka} which are all flaring nova events. The third section are two sources gathered manually and the final section contains CVs from \citet{Deanne}.The distance references are as follows: [1]~\citet{TCrB_distance}; [2]~\citet{CVdistances}; [3]~\citet{RSOph_distance}; [4]~\citet{NovaSco_o}; [5]~\citet{TPyx_distance}; [6]~\citet{V1500Cyg_distance}; [7]~\citet{V1723Aql_r}; [8]~\citet{V1974Cyg_distance}; [9]~\citet{gaiaCVdistances}; [10]~\citet{Deanne}. Below the table are notes on individual sources regarding the circumstance of the $F_{\mathrm{o}}$ and $F_{\mathrm{r}}$ measurements used. If no note is present for a source it can be assumed to be in quiescence.}
	\label{table:CVs}
	\begin{tabular}{ccccc} 
		\hline
		Object Name & CV Subclass & $F_{\mathrm{o}}$ ref. & $F_{\mathrm{r}}$ ref. & d (pc)\\
		\hline
		T CrB & Nova &\citet{Gudel} & \citet{Gudel} & 1,063.8 [1]\\
		V* RT Ser & Nova & \citet{Gudel} & \citet{Gudel} & $\geq$ 1,200 [2]\\
		RS Oph$^a$ & Nova & \citet{Gudel} & \citet{Gudel} & 1,400 [3]\\
		V 4074 Sgr & Nova & \citet{Gudel} & \citet{Gudel} & - \\
		HM Sge & Nova & \citet{Gudel} & \citet{Gudel} & $\geq$1,500 [2]\\
		PU Vul & Nova & \citet{Gudel} & \citet{Gudel} & -\\
		AG Peg & Nova & \citet{Gudel} & \citet{Gudel} & -\\
		\hline
		Nova~Sco~2012 & Nova & \citet{NovaSco_o} & \citet{NovaSco_r} & 3,700--4,300 [4]\\	
		T~Pyx & Nova & \citet{TPyx} & \citet{TPyx} & 4800 [5]\\
		V1500~Cyg & Nova & AAVSO & \citet{V1500Cyg_r} & 1950 [6] \\
		V1723~Aql & Nova & \citet{V1723Aql_o} & \citet{V1723Aql_r} & 5000 [7] \\
		V1974~Cyg & Nova & AAVSO & \citet{V1974Cyg_r} & 1830 [8]\\
		\hline
		SS Cyg & Dwarf Nova & \citet{SSCyg} & \citet{SSCyg} & 117.1 [9]\\
		V3885 Sgr & Dwarf Nova & AAVSO & \citet{V3885_r} & 136.0 [9]\\
		\hline
		RW Sex & Novalike & \citet{Deanne}/AAVSO & \citet{Deanne} & 150.0 [10]\\
		V603 Aql & Novalike & \citet{Deanne}/AAVSO & \citet{Deanne} & 328.3 [9]\\
		TT Ari$^b$ & Novalike & \citet{Deanne}/AAVSO & \citet{Deanne} & 228.8 [9]\\
		\hline
	\end{tabular}
		\begin{flushleft}
		$^a$ Data for RS~Oph was also obtained from different sources for the dynamic sample (Table~\ref{table:dynamic}).\\
		$^b$ Two measurements were included in the sample of TT~Ari - at radio outburst and quiescence.\\ 
		\textbf{Notes on the circumstance of the $F_{\mathrm{o}}$ and $F_{\mathrm{r}}$ measurements obtained for specific sources (refer to main text in Section~\ref{sec:CVs} for those without a note):}\\
		\textbf{RS~Oph (Dynamic):} RS~Oph is coverage of a flare that begun on 2006~February~12 and was first detected in the radio 4~d after the optical peak (day 0). 
		\textbf{Nova Sco 2012:} The $F_{\mathrm{o}}$ and $F_{\mathrm{r}}$ measurements were recorded 35~d after the discovery of the nova as an optical transient on 2012~May~22. This was the first radio detection, and it was made after the optical maximum.\\
		\textbf{T Pyx:} The $F_{\mathrm{o}}$ and $F_{\mathrm{r}}$ measurements used are 66~d after the optical transient discovery on 2011~April~14. This was the first radio detection at 5~GHz. The radio peak occurred $\sim$250~d post discovery.\\
		\textbf{V1500 Cyg:} The $F_{\mathrm{o}}$ and $F_{\mathrm{r}}$ measurements used are 27~d after the optical transient discovery on 1975~August~28. This was the first radio detection at 8~GHz. The radio peak occurred $\sim$106~d post discovery.\\
		\textbf{V1723 Cyg:} The $F_{\mathrm{o}}$ and $F_{\mathrm{r}}$ measurements used are 3~d after the optical transient discovery on 2010~September~11. This was the first radio detection at 5~GHz. The radio peak occurred $\sim$57~d post discovery. \\
		\textbf{V1974 Cyg:} The $F_{\mathrm{o}}$ and $F_{\mathrm{r}}$ measurements used are 21~d after the optical transient discovery on 1992~February~19. This was the first radio detection at 8.4~GHz. The radio peak occurred $\sim$250~d post discovery.\\
		\textbf{SS Cyg:} The $F_{\mathrm{o}}$ and $F_{\mathrm{r}}$ measurements used are 1.3~d after the optical flare was detected on 2007~April~13. This was at the radio peak of the observations which then declined over the next 20~d back to below the detection threshold.\\
		\textbf{V3885 Sgr:} Targeted radio observation with a single detection, matched with optical 38~d later as assumed to be in a quiescent state.\\
		\textbf{TT Ari:} Two measurements are included here as one of the targeted radio observations performed by the authors happened to detect a radio flare showing an increase in peak flux density of 200~$\upmu$Jy from the previous detection at 40~$\upmu$Jy.
		\end{flushleft}
\end{table*}

\begin{table*}
	\centering
	\caption{The XRB sources that are included in the sample. The sample was gathered with the use of two catalogues: the Catalogue of high-mass X-ray binaries in the Galaxy \citep[4$^{\textrm{th}}$~edition;][]{HMXBcat}, that accounts for the first section of the table, and the Catalogue of low-mass X-ray binaries in the Galaxy, LMC, and SMC \citep[4$^{\textrm{th}}$~edition;][]{LMXBcat} which accounts for the second section of the table. If the distance does not have an associated reference then it was collected from the respective catalouge. In other cases the distance references are as follows: [1]~\citet{CirX1distance}; [2]~\citet{xte1550distance1}; [3]~\citet{xte1550distance2}; [4]~\citet{v404distance}. Below the table are notes on the circumstance of the collected $F_{\mathrm{o}}$ and $F_{\mathrm{r}}$ measurements for each individual source.}
	\label{table:x-ray}
	\resizebox{\textwidth}{!}{\begin{tabular}{ccccc} 
		\hline
		Object Name & Subclass& $F_{\mathrm{o}}$ ref. & $F_{\mathrm{r}}$ ref. & d (kpc)\\
		\hline
            LS I +61 303 &          HMXB & \citet{lsi61303_o} &\citet{lsi61303_r}  & 2.4 \\
           XTE J0421+560 &          HMXB & \citet{xtej0421} & \citet{xtej0421} &    1 -- 5  \\
         SAX J1819.3-2525 &          HMXB & \citet{saxj1819} & \citet{saxj1819} &  7 -- 12\\
         RX J1826.2-1450 &          HMXB & \citet{rxj1826} & \citet{rxj1826} & 2.5 \\
     			SS433 &          HMXB & \citet{ss433_o} & \citet{ss433_r} &  5.5  \\
              4U 1956+35 &          HMXB & \citet{4u1956_o} & \citet{4u1956_r} &  2.2  \\
\hline
         IGR J00291+5934 &          LMXB & \citet{igrj00291_o} & \citet{igrj00291_r} & 2.6 -- 3.6 \\
            GRO J0422+32 &          LMXB & \citet{groj0422} & \citet{groj0422} &  2.4\\
             4U 0614+091 &          LMXB & \citet{4u0614} & \citet{4u0614} & < 4 \\
              1A 0620-00 &          LMXB & \citet{a0620_o} & \citet{a0620_r} & 1.2 \\
           XTE J0929-314 &          LMXB & \citet{xtej0929_o} & \citet{xtej0929_r} & 5 -- 15 \\
              GS 1124-684  &          LMXB & \citet{grs1124_o} & \citet{grs1124_r} & 5.9 \\
              GS 1354-64 &          LMXB & \citet{gs1354} & \citet{gs1354} &  $\geq 27$\\
              4U 1456-32 &          LMXB & \citet{4u1456_o} & \citet{4u1456_r} & 1.2 \\
             3A 1516-569 &  \ \ LMXB$^a$ & \citet{3a1516} & \citet{3a1516} & 8.4 -- 10.2 [1] \\
              4U 1543-47 &          LMXB & \citet{4u1543_o} & \citet{4u1543_r} & 7.5 \\
           XTE J1550-564 &          LMXB & \citet{xtej1550_o} & \citet{xtej1550_r} &  2.5 -- 6 [2,3]\\
             GRO~J1655-40& LMXB & \citet{groj1655_r} (SMARTS$^b$) & \citet{J165540_r_early,groj1655_r} & 3.2 \\
                    GX~339-4 & LMXB & \citet{Buxton} & \citet{Corbel}  & 8.0 [2]  \\
            GRO J1719-24 &          LMXB & \citet{groj1719} & \citet{groj1719} &  $\sim2.4$  \\
            GRS 1739-278 &          LMXB & \citet{grs1739_o} & \citet{grs1739_r} &  8.5  \\
      Swift J1753.5-0127 &          LMXB & \citet{swiftj1753_o} & \citet{swiftj1753_r} &  $\sim6$  \\
        SAX J1808.4-3658 &          LMXB & \citet{saxj1808_o} & \citet{saxj1808_r} &  2-5 -- 3.6  \\
           XTE J1859+226 &          LMXB & \citet{xtej1859_o} & \citet{xtej1859_r} (GBI) &  7.6 \\
		Aql~X-1 & LMXB & SMARTS$^a$ & \citet{aqlx1_r} & 5.2 \\
             V404~Cygni & LMXB & \citet{Casares} & \citet{Han} &  2.4 [4] \\
		\hline
	\end{tabular}}
		\begin{flushleft}
{\scriptsize
		$^a$ The exact nature of the companion of 3A 1516-569 (Cir~X-1) is not well known (e.g. \citet{CirX1_doubt}), so while it is in the LMXB catalogue, it is classed as a HMXB in other sources.\\
		$^b$ Optical and IR monitoring of X-ray binaries using SMARTS \citep{Buxton}.\\
		\textbf{Notes on the circumstance of the $F_{\mathrm{o}}$ and $F_{\mathrm{r}}$ measurements obtained for each source:}\\
		\textbf{LS~I~+61 303:} Measurements are averages of long-term monitoring ($\sim2$~months) with no flaring.\\
\textbf{XTE J0421+560:} 4~d after the detection as an X-ray transient by the Rossi X-ray Timing Explorer (RXTE) on 1998~March~31. Radio peak is used.\\
         \textbf{SAX J1819.3-2525:} Less then 1~d after the detection of an X-ray flare by RXTE on 1999 September 15. Radio peak is used.\\
         \textbf{RX J1826.2-1450:} Targeted radio detection. $F_{\mathrm{o}}$ measured 26~d prior. Low level variabilty at time of measurements (10~per~cent radio, 0.05~mag optical).\\
    \textbf{SS433:} Radio flare detected on 2014 September 23. $F_{\mathrm{o}}$ recorded previous day ($<0.5$~d).\\
              \textbf{4U 1956+35:} Measurements from two separate long-term monitoring campaings that overlapped in 1977. No flaring seen at time of used measurements.  \\
         \textbf{IGR J00291+5934:} $F_{\mathrm{o}}$ and $F_{\mathrm{r}}$ measurements taken 2~d and 4.5~d, respectively, after X-ray transient detection by \textit{INTEGRAL} on 2004 December 02. \\
            \textbf{GRO J0422+32:} Approximately 2~weeks after X-ray transient detection by the BATSE experiment on 1992 August 05.\\
             \textbf{4U 0614+091:} Multiwavelength campaign to measure the spectrum. $F_{\mathrm{o}}$ and $F_{\mathrm{r}}$ are 2.5~d apart with no major flaring event.\\
              \textbf{1A 0620-00:} Measurements approximately 2~weeks after the detection as an X-ray transient on 1975 August 03.\\
           \textbf{XTE J0929-314:} $F_{\mathrm{o}}$ and $F_{\mathrm{r}}$ measurements taken 4~d after X-ray transient detection by RXTE on 2002 April 30. \\
              \textbf{GS 1124-684:} $F_{\mathrm{o}}$ and $F_{\mathrm{r}}$ measurements taken 10~d after the discovery as an X-ray transient on 1991 January 08 by the GRANAT satellite. Radio peak is used.\\
              \textbf{GS 1354-64:} $F_{\mathrm{o}}$ and $F_{\mathrm{r}}$ measurements taken following the detection of a hard-state outburst in 1997 November, mostly in the decay phase. \\
              \textbf{4U 1456-32:} $F_{\mathrm{o}}$ and $F_{\mathrm{r}}$ measurements taken 10~d and 14~d, respectively, after X-ray transient detection by the Ariel 5 all-sky monitor on 1979~May~11. \\
             \textbf{3A 1516-569:} Simultaneous $F_{\mathrm{o}}$ and $F_{\mathrm{r}}$ measurements during a flare on 1977 May 12, very close to radio peak. \\
              \textbf{4U 1543-47:} $F_{\mathrm{o}}$ and $F_{\mathrm{r}}$ measurements taken 1d after an X-ray outburst was detected by RXTE on 2002 June 16. \\
           \textbf{XTE J1550-564:} Measurements used are at the radio peak, approximately 2~weeks after discovery as an X-ray transient by RXTE on 1998 September 07.\\
             \textbf{GRO~J1655-40:} Measurements taken at the radio peak which was approximately 1~month after the RXTE detection of an outburst on 2005 February 17.\\
                    \textbf{GX~339-4:} Measurements taken on 2011 Feb 06 during the decay of an outburst early in the previous year 2010.\\
            \textbf{GRO J1719-24:} $F_{\mathrm{r}}$ and $F_{\mathrm{o}}$ measurements taken 8~d and 10~d, respectively, after X-ray transient detection by BATSE on 1993~September~25. \\
            \textbf{GRS 1739-278:} $F_{\mathrm{o}}$ and $F_{\mathrm{r}}$ measurements taken approximately 1~month after X-ray transient detection on 1996~March~18.\\
      \textbf{Swift J1753.5-0127:} $F_{\mathrm{o}}$ and $F_{\mathrm{r}}$ measurements taken 2~d and 3~d, respectively, after new gamma source detection by Swift BAT on 2005~June~30. \\
        \textbf{SAX J1808.4-3658:} $F_{\mathrm{o}}$ and $F_{\mathrm{r}}$ measurements taken 18~d after an outburst was detected by RXTE on 1998~April~09.\\
           \textbf{XTE J1859+226:} $F_{\mathrm{o}}$ and $F_{\mathrm{r}}$ measurements taken 7~d after X-ray transient detection by RXTE on 1999~October~07. This was the radio peak.\\ 
		\textbf{Aql~X-1:} $F_{\mathrm{o}}$ and $F_{\mathrm{r}}$ measurements taken 15~d after the detection of an outburst in the optical on 2013~June~4. \\
             \textbf{V404~Cygni:} $F_{\mathrm{o}}$ and $F_{\mathrm{r}}$ measurements taken 9~d after X-ray outburst detection on 1989~May~21. This was the radio peak.
}
		\end{flushleft}
\end{table*}

\begin{table*}
	\centering
	\caption{The SN events that are included in this work. The age column signifies the approximate time elapsed since the accepted explosion date of the respective SN. The sample is based upon the list of radio SNe compiled by \citet{Romero}. The distance references are as follows: [1]~\citet{SNunifiedcatalogue} and [2]~\citet{OpenSNCatalog}. }
	\label{table:SNe}
	\begin{tabular}{cccccc} 
		\hline
		SN Name & SN Type & $F_{\mathrm{o}}$ ref. & $F_{\mathrm{r}}$ ref. & Age (d) & d$_L$ (Mpc)\\
		\hline
		SN 1980K & IIb & \citet{Buta} & \citet{Weiler86} & 37 & 4.7 [1]\\
		SN 1984L & Ib & \citet{Tsvetkov} & \citet{Panagia} & 36 & 18.0 [1]\\
		SN 1988Z & IIn & \citet{Turatto} & \citet{Williams} & 743 & 70.0 [1]\\
		SN 1990B & Ic & \citet{Clocchiatti} & \citet{vanDyk93} & 61 & 24.0 [1]\\
		SN 1993J & II & \citet{Richmond} & \citet{Weiler} & 11 & 2.9 [1]\\
		SN 1994I & Ic & \citet{SN1994I_o} & \citet{SN1994I_r} & 5 & 6.1 [1]\\
		SN 1996cb & IIb & \citet{Qiu} & \citet{vanDyk} & 9 & 3.8 [1]\\
		SN 1998bw & Ic & \citet{SN1998bw_track} & \citet{SN1998bw_r} & 16 & 37.9 [2]\\
		SN 1999em & IIP & \citet{Leonard} & \citet{Pooley} & 34 & 7.7 [1]\\
		SN 2001ig & IIb & \citet{Bembrick} & \citet{Ryder} & 12 & 9.0 [1]\\
		SN 2002ap & Ic & \citet{Gal-Yam} & \citet{Berger} & 4 & 3.4 [1]\\
		SN 2002hh & II & \citet{Pozzo} & \citet{Stockdale02} & 17 & 4.7 [1]\\
		SN 2003L & Ic & VSNET$^a$ & \citet{Soderberg05} & 32 & 95.3 [2]\\
		SN 2003bg & IIb & \citet{Hamuy} & \citet{Soderberg} & 10 & 20.3 [2]\\
		SN 2004dj & IIP & \citet{Schmeer} & \citet{Stockdale} & 23 & 3.5 [1]\\
		SN 2004dk & Ib & \citet{Drout} & \citet{Wellons} & 8 & 20.0 [1]\\
		SN 2004et & II & \citet{Misra} & \citet{Stockdale04} & 14 & 4.7 [1]\\
		SN 2004gq & Ib & \citet{Drout} & \citet{Wellons} & 8 & 16.0 [1]\\
		SN 2007bg & Ic & \citet{SN2007bg_o} & \citet{SN2007bg_r} & 19 & 154.0 [2]\\
		SN 2007gr & Ic & \citet{SN2007gr_o} & \citet{SN2007gr_r} & 5 & 10.0 [1]\\
		SN 2007uy & Ib & \citet{SN2007uy} & \citet{SN2007uy} & 9 & 26.0 [1]\\
		SN 2008D & Ib & \citet{SN2008D_o} & \citet{SN2008D_r} & 19 & 26.0 [1]\\
		SN 2008ax & IIb & \citet{Roming} & \citet{Roming} & 8 & 5.1 [1]\\
		SN 2009bb & Ic & \citet{SN2009bb_o} & \citet{SN2009bb_r} & 20 & 46.4 [2]\\
		SN 2011hs & IIb & \citet{SN2011hs} & \citet{SN2011hs} & 12 & 25.3 [2]\\
		SN 2011dh & IIb & \citet{Arcavi} & \citet{Horesh} & 3 & 7.3 [2]\\
		\hline
	\end{tabular}
		\begin{flushleft}
		$^a$ Variable Star Network \citep[VSNET;][]{VSNET}\\
		\end{flushleft}
\end{table*}

\begin{table*}
	\centering
	\caption{GRB data included in this work which is based upon the list of radio GRB-afterglows compiled by \citet{Chandra}. All are long GRBs apart from the two denoted. The references for the optical and radio measurements along with the age column that signifies the approximate time elapsed since the detection of the burst. Redshifts, $z$, are also given where available. If no reference for $z$ is stated then the value is taken from one of the measurement references, the stated references are as follows:  [1]~\citet{GRB980329redshift}; [2]~\citet{GRB980703redshift}; [3]~\citet{GRB990123redshift}; [4]~\citet{GRB990510redshift}; [5]~\citet{GRB991216redshift}; [6]~\citet{GRB000301Credshift}; [7]~\citet{GRB000418redshift}; [8]~\citet{GRB010222redshift}; [9]~\citet{GRB020124redshift}; [10]~\citet{GRB020813redshift}; [11]~\citet{030329_redshift}; [12]~\citet{GRB050315redshift}; [13]~\citet{GRB050416aredshift}; [14]~\citet{GRB050603redshift}; [15]~\citet{GRB050820Aredshift}; [16]~\citet{GRB051109redshift}; [17]~\citet{GRB060218redshift}; [18]~\citet{GRB061121redshift}; [19]~\citet{GRB071010Bredhift}; [20]~\citet{GRB071020redshift}; [21]~\citet{GRB080319Bredshift}; [22]~\citet{GRB090313redshift}; [23]~\citet{GRB090323redshift}; [24]~\citet{GRB090424redshift}; [25]~\citet{GRB090618redshift}; [26]~\citet{GRB090902Bredshift}; [27]~\citet{GRB091020redshift}; [28]~\citet{100418A_redshift}; [29]~\citet{GRB100814Aredshift}; [30]~\citet{GRB100901Aredshift}.}
	\label{table:GRBs}
	\resizebox{\textwidth}{!}{\begin{tabular}{ccccc} 
		\hline
		GRB Name & $F_{\mathrm{o}}$ ref. & $F_{\mathrm{r}}$ ref. & Age (d) & $z$ \\
		\hline
		GRB 970508 & \citet{GRB970508_o} & \citet{GRB970508_r} & 5 & 0.835\\
		GRB 980329 & \citet{GRB980329_o} & \citet{GRB980329_r} & $< 1$ & $\approx2$, or $3\lesssim z \lesssim 5$ [1] \\
		GRB 980519 & \citet{GRB980519_o} & \citet{GRB980519_r} & 3 & - \\
		GRB 980703 & \citet{GRB980703_o} & \citet{GRB980703_r} & 1 & 0.966 [2]\\
		GRB 990123 & \citet{GRB990123_o} & \citet{GRB990123_r} & 1 & 1.600 [3]\\
		GRB 990510 & \citet{GRB990510} & \citet{GRB990510} & 1 & 1.619 [4] \\
		GRB 991208 & \citet{GRB991208_o} & \citet{GRB991208_r} & 1 & 0.706\\
		GRB 991216 & \citet{GRB991216_o} & \citet{GRB991216_r} & 2 & 1.02 [5] \\
		GRB 000131 & \citet{GRB000131_o} & \citet{GRB000131_r} & 8 & 4.500\\
		GRB 000301C & \citet{GRB000301C_o} & \citet{GRB000301C_r} & 4 & 2.034 [6] \\
		GRB 000418 & \citet{GRB000418} & \citet{GRB000418} & 11 & 1.118 [7] \\
		GRB 000911 & \citet{GRB000911} & \citet{GRB000911} & 4 & 1.059\\
		GRB 000926 & \citet{GRB000926_o} & \citet{GRB000926_r} & 3 & 2.038\\
		GRB 010222 & \citet{GRB010222_o} & RADW$^a$ & 1 & 1.477 [8] \\
		GRB 010921 & \citet{GRB010921} & \citet{GRB010921} & 26 & 0.451\\
		GRB 011121 & \citet{GRB011121_o} & \citet{GRB011121_r} & 4 & 0.362\\
		GRB 020124 & \citet{GRB020124} & \citet{GRB020124} & 2 & 3.200 [9]\\
		GRB 020405 & \citet{GRB020405_o} & \citet{GRB020405_r}& 1 & 0.691\\
		GRB 020813 & \citet{GRB020813_o} & \citet{GRB020813_r}& 1 & 1.255 [10]\\
		GRB 030115 & \citet{GRB030115_o} & \citet{GRB030115_r}& 2 & $\sim$2.500\\
		GRB 030329 & \citet{GRB030329_o} & \citet{GRB030329_r}& $ < 1$ & 0.169 [11] \\
		GRB 050315 & \citet{GRB050315_o} & \citet{GRB050315_r}& 1 & 1.949 [12] \\
		GRB 050401 & \citet{GRB050401_o} & \citet{GRB050401_r}& 5 & 2.899\\
		GRB 050525A & \citet{GRB050525A_o} & \citet{GRB050525A_r}$^{b}$& 3 & 0.606\\
		GRB 050603 & \citet{GRB050603_o} & \citet{GRB050603_r}& $< 1$ & 2.821 [14]\\
		GRB 050724$^{c}$ & \citet{GRB050724_o} & \citet{GRB050724_r}& $< 1$ & 0.257\\
		GRB 050820A & \citet{GRB050820A} & \citet{GRB050820A}& $< 1$ & 2.615 [15]\\
		GRB 051109A & \citet{GRB051109A_o} & \citet{GRB051109A_r}& 1 & 2.346 [16]\\
		GRB 051221A$^{c}$ & \citet{GRB051221A} & \citet{GRB051221A}& 1 & 0.546\\
		GRB 060218 & \citet{GRB060218_o} & \citet{GRB060218_r}& 2 & 0.033 [17]\\
		GRB 061121 & \citet{GRB061121_o} & \citet{GRB061121_r}& 1 & 1.314 [18]\\
		GRB 070125 & \citet{GRB070125} & \citet{GRB070125}& 2 & 1.547\\
		GRB 071003 & \citet{GRB071003} & \citet{GRB071003}& 2 & 1.604\\
		GRB 071010B & \citet{GRB071010B_o} & \citet{GRB071010B_r}& 3 & 0.947 [19]\\
		GRB 071020 & \citet{GRB071020_o} & \citet{GRB071020_r}& 2 & 2.142 [20]\\
		GRB 080319B & \citet{GRB080319B_o} & \citet{GRB080319B_r} & 2 & 0.937 [21] \\
		GRB 080603A & \citet{GRB080603A_o} & \citet{GRB080603A_r}& 2 & 1.687\\
		GRB 080810 & \citet{GRB080810_o} & \citet{GRB080810_r}& 4 & 3.355\\
		GRB 081203B & \citet{GRB081203B_o} & \citet{GRB081203B_r}& 5 & - \\
		GRB 090313 & \citet{GRB090313} & \citet{GRB090313}& 6 & 3.375 [22]\\
		GRB 090323 & \citet{GRB090323_o} & \citet{GRB090323_r}& 5 & 3.570 [23]\\
		GRB 090424 & \citet{GRB090424_o} & \citet{GRB090424_r}& 2 & 0.544 [24]\\
		GRB 090618 & \citet{GRB090618_o} & \citet{GRB090618_r}& 1 & 0.540 [25]\\
		GRB 090902B & \citet{GRB090902B_o} & \citet{GRB090902B_r}& 2 & 1.822 [26]\\
		GRB 091020 & \citet{GRB091020_o} & \citet{GRB091020_r}& 1 & 1.71 [27]\\
		GRB 100418A & \citet{GRB100418A_o} & \citet{GRB100418A_r}& 2 & 0.624 [28] \\
		GRB 100814A & \citet{GRB100814A_o} & \citet{GRB100814A_r}& 4 & 1.440 [29]\\
		GRB 100901A & \citet{GRB100901A_o} & \citet{GRB100901A_r}& 3 & 1.408 [30] \\
		\hline
	\end{tabular}}
	\begin{flushleft}
	$^a$ Measurement taken from the `Radio Afterglow Data Website' (RADW; \url{http://www.aoc.nrao.edu/~dfrail/allgrb_table.shtml}).\\
	$^b$ Closer measurement in time taken from RADW.\\
         $^c$ Short GRB.\\
	\end{flushleft}
\end{table*}

\bsp	
\label{lastpage}
\end{document}